\documentclass[a4paper,fleqn,usenatbib]{mnras}

\usepackage{txfonts}

\usepackage{mathptmx}

\usepackage[T1]{fontenc}
\usepackage{ae,aecompl}

\usepackage{graphicx}	
\usepackage{amssymb}	
\usepackage{longtable}
\usepackage{multicol}

\title[]{An all-sky Support Vector Machine selection of WISE YSO Candidates}

\author[G. Marton et al.]{
G. Marton,$^{1}$\thanks{E-mail: marton.gabor@csfk.mta.hu}
L. V. Toth,$^{2}$
R. Paladini,$^{3}$
M. Kun,$^{1}$
S. Zahorecz,$^{2,4}$
\newauthor
P. McGehee,$^{3}$
Cs. Kiss$^{1}$
\\
$^{1}$Konkoly Observatory, Research Centre for Astronomy and Earth Sciences, Hungarian Academy of Sciences, H-1121 Budapest\\
$^{2}$Lor\'and E\"otv\"os University, Department of Astronomy, P\'azm\'any P.s. 1/a, H-1117 Budapest, Hungary\\
$^{3}$Infrared Processing Analysis Center, California Institute of Technology, 770 South Wilson Ave., Pasadena, CA 91125, USA\\
$^{4}$European Southern Observatory, Karl-Schwarzschild-Str. 2, 85748, Garching bei M\"unchen, Germany\\
}

\date{Accepted XXX. Received YYY; in original form ZZZ}

\pubyear{2015}

\begin{document}
\label{firstpage}
\pagerange{\pageref{firstpage}--\pageref{lastpage}}
\maketitle

\begin{abstract}
We explored the AllWISE catalogue of the Wide-field Infrared Survey Explorer mission and identified Young Stellar Object candidates. Reliable 2MASS and WISE photometric data combined with Planck dust opacity values were used to build our dataset and to find the best classification scheme. A sophisticated statistical method, the Support Vector Machine (SVM) is used to analyse the multi-dimensional data space and to remove source types identified as contaminants (extragalactic sources, main sequence stars, evolved stars and sources related to the interstellar medium). Objects listed in the SIMBAD database are used to identify the already known sources and to train our method. A new all-sky selection of 133,980 Class I/II YSO candidates is presented. The estimated contamination was found to be well below 1\% based on comparison with our SIMBAD training set. We also compare our results to that of existing methods and catalogues. The SVM selection process successfully identified $>90\%$ of the Class I/II YSOs based on comparison with photometric and spectroscopic YSO catalogues. Our conclusion is that by using the SVM, our classification is able to identify more known YSOs of the training sample than other methods based on colour-colour and magnitude-colour selection. The distribution of the YSO candidates well correlates with that of the Planck Galactic Cold Clumps in the Taurus--Auriga--Perseus--California region.
\end{abstract}




\section{Introduction}

\begin{figure}
 \includegraphics[width=0.5\textwidth]{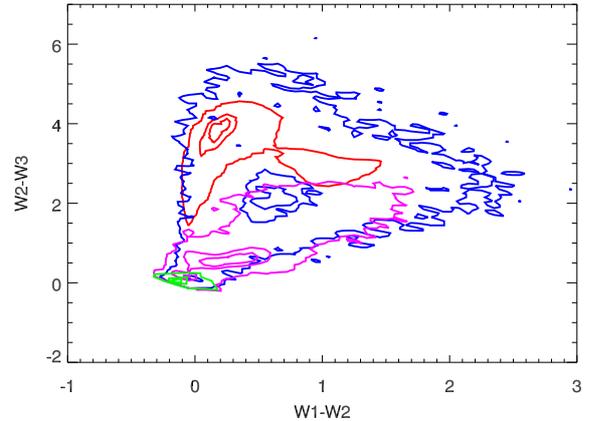}
\caption{Object types identified with SIMBAD on the W1$-$W2, W2$-$W3 plane. The figure demonstrates that different object types are overlapping, boundaries between them are non-linear. Contour lines show the 5\%, 50\% and 75\% of the maximal surface density of extragalactic sources (red), field stars (green), evolved stars (magenta) and YSOs (blue). The surface density of different types was calculated in bins of 0.1 mag.}
\label{w1w2w2w3}
\end{figure}

The amount of data collected by infrared (IR) satellites and observatories has been continuously increasing over the past three decades. The evolution of the detectors allowed us to explore the interstellar medium (ISM) and embedded objects in more and more detail. IRAS \citep{neugebauer1984} provided $\sim350,000$ objects with flux above 0.5 Jy at 12 $\mu m$. Recently, based on observations of the WISE \citep{wright2010} infrared satellite, more than 700 million sources with $>5\sigma$ accuracy above 1 mJy were catalogued in the AllWISE Data Release \citep{cutri2013}. Infrared luminous objects cover a broad spectrum of object types. Extragalactic sources, especially galaxies with ongoing star formation or AGNs, show similar spectral energy distribution (SED) to that of the Young Stellar Objects (YSOs). Evolved stars eject dust into their outer envelopes, which has infrared colours analogous to the dust surrounding YSOs in their early evolutionary stages. In this work we identify the AllWISE sources by searching for a close counterpart in the SIMBAD database, using a 5$^{\prime\prime}$ radius. The closest SIMBAD source within this radius is associated with the AllWISE object. Figure \ref{w1w2w2w3} illustrates the surface density of the known extragalactic sources, main sequence stars, evolved stars and YSOs in the WISE W1$-$W2, W2$-$W3 colour--colour plane and it shows that the different object types have highly overlapping WISE colours. Linear methods would fail to separate the different object types and result in samples with high percentage of contamination, so the separation requires special attention.

The complexity of the observable properties makes the object classification a fundamental and challenging problem. However, the commonly used schemes do not take advantage of all the available information. Sources are often identified on colour--colour and colour--magnitude diagrams: \citet{gutermuth2008} characterized the Spitzer \citep{werner2004} IRAC \citep{fazio2004} colour and magnitude properties of proto-- and pre--main sequence stars in NGC 1333, and completed the dataset with MIPS \citep{rieke2004} and J, H and K$_s$ 2MASS \citep{cutri2003,skrutskie2006} data. This method was then extended by \citet{gutermuth2009} to several star forming clouds, located within 1 kpc of the Sun. \citet{rebull2010} also used IRAC and MIPS colours together with 2MASS data to identify YSOs in the Taurus Molecular Cloud (TMC). They note that the method does not seem to successfully weed out all the galaxies. \citet{harvey2007} set criteria for YSO identification on large scale maps of star forming regions, with a primary goal of mitigation of extragalactic contamination. This kind of large scale classification is problematic because of the wide variation of extinction. \citet{rebull2011} also identified YSO candidates in the TMC by using WISE photometry, as described in \citet{koenig2012}. By using far-IR (four bands between 65 and 160 $\mu$m) AKARI \citep{murakami2007} FIS \citep{kawada2007} colours and flux densities, \citet{pollo2010} successfully separated the sources of the AKARI Bright Source Catalogue \citep[BSC]{yamamura2010} in low-extinction regions by classifying them as either extragalactic sources or Milky Way stars. Based on a combination of far--IR AKARI and mid--IR WISE data, \citet{toth2014} used Quadratic Discriminant Analysis to identify YSO candidates. Their comparison to the known YSOs of the SIMBAD database showed that 90\% of the training sample YSOs were successfully reclassified as YSO candidates, while the fraction of known contaminants remained $<$10\%. 

In this paper we built a multi--dimensional dataset containing near--IR 2MASS and mid--IR WISE data and we apply the Support Vector Machine (SVM) method to identify potential YSO candidates. Our goal is to create a catalogue of carefully selected YSO candidates that can be used for statistical studies, and that can also provide a list of potential targets for future follow-up observations. We show that SVM \citep{vapnik1995}, which is a commonly used tool in pattern recognition and in multi--dimensional classification, is able to identify higher fraction of the training sample YSOs than the regularly used polygonal selections on colour--colour and colour--magnitude planes. We have to note that, due to the lack of spectroscopic data, our training samples are based on SIMBAD identifications. Therefore, most of our comparisons are also estimates based on SIMBAD. We identify extragalactic contaminants, galactic field stars, galactic evolved stars and other galactic contamination. As a result, YSO candidates are identified and an attempt is made to separate them based on their evolutionary stages. Verification of our method and comparison to existing methods of YSO selection has been done. We also investigate, in a well-known star forming region (i.e. the Taurus--Auriga--Perseus--California molecular cloud), the potential spatial correlation between the  candidate YSOs and the Cold Clumps listed in the Planck Catalogue \citep{planck2015}.

\section{Data and Method}

\subsection{Data}\label{data}

To search for the YSO candidates we used the AllWISE Data Release \citep{cutri2013}, which is an improved version of the WISE All-Sky Data Release \citep{cutri2012}. The AllWISE catalog contains information on 747,634,026 sources. We used not only brightness values in all 4 WISE passbands (3.6, 4.6, 12 and 22 $\mu m$), but  also the extended source flag (\textit{ext}), which indicates whether or not the morphology of a source is consistent with the WISE point spread function, and also if the source is associated with or superimposed on a previously known extended object from the 2MASS Extended Source Catalog (XSC). We also used 2MASS J, H, K$_s$ magnitudes, which are provided in the AllWISE catalog, based on 2MASS PSC \citep{cutri2003} associations. Instead of the whole AllWISE catalogue, we used only those sources that matched the following criteria: (1) SNR$>$3 in all four WISE bands and (2) 2MASS J, H, K$_s$ magnitudes are available with photometric errors lower than 0.1 mag. Applying these criteria resulted in 8,956,636 sources. These form our initial sample, which hereafter we call the \textit{W0} sample. 

At the position of each source we estimated the dust optical depth ($\tau$) by using the 353GHz R1.2 Planck dust opacity maps \citep{planck2014}. This operation allowed us to include the effect of interstellar reddening in our analysis.

\subsection{Support Vector Machines}\label{svm}
For classification and pattern recognition in multi--dimensional data, one can use several statistical methods. We used the Support Vector Machine (SVM), a class of supervised learning algorithm, developed by Vladimir Vapnik \citep{vapnik1995} as an extension to non-linear models of the generalized portrait algorithm. SVM calculates decision planes between different known classes of objects and applies the decision planes to objects of unknown classes. These unknown objects are classified based on their position in the multi--dimensional parameter space with respect to the separation boundaries. A more detailed description of the method can be found in \citet{malek2013}. Various statistical methods are usually among the major ingredients of professional statistical software packages. We used the $R$ implementation of SVM in our work.

SVM is a supervised learning algorithm, therefore it needs a training set, which is used to determine the boundaries in the parameter space between the different object types. To find out the object types of our sources, we searched the SIMBAD database for counterparts within $5^{\prime\prime}$ radius. If more than one object was located within this radius, then the type of the closest entry was used. More information about the SIMBAD object identification can be found in \citet{ochsenbein1992}. Following this strategy, we were able to identify 890,552 sources of the \textit{W0} sample. We are aware that individual source identifications in SIMBAD might not be fully reliable. However, for our purposes, we only need training samples that are statistically reliable. The number of sources found per object type are listed in the \textit{W0} column of Table \ref{simbadcomparison}. Unless noted otherwise, in the remaining of this paper the expression "known object" always refers to a SIMBAD identification. The average colours and magnitudes for each SIMBAD object type in our \textit{W0} sample are listed in Tables \ref{simbadcolours1}--\ref{simbadcolours9}.

Because the number of known sources from the different object types was very inhomogeneous, e.g. much more known extragalactic objects and field stars were found than YSOs, we checked if the classification is affected by the number of sources used in the training samples. To this end, we considered two cases: a) we used all objects of given object types, b) we used a maximum number of 1000 sources of each object types. We found that the number of false-positives and false-negatives differs by less than 1\% between case a) and b). Therefore in our classification scheme the number of objects used in the training sample was always limited to 1,000, allowing us to speed up our classification process. 

\section{Classification steps and results}\label{scheme}
\begin{figure*}
 \includegraphics[width=1.0\textwidth,trim=0cm 4cm 0cm 0cm]{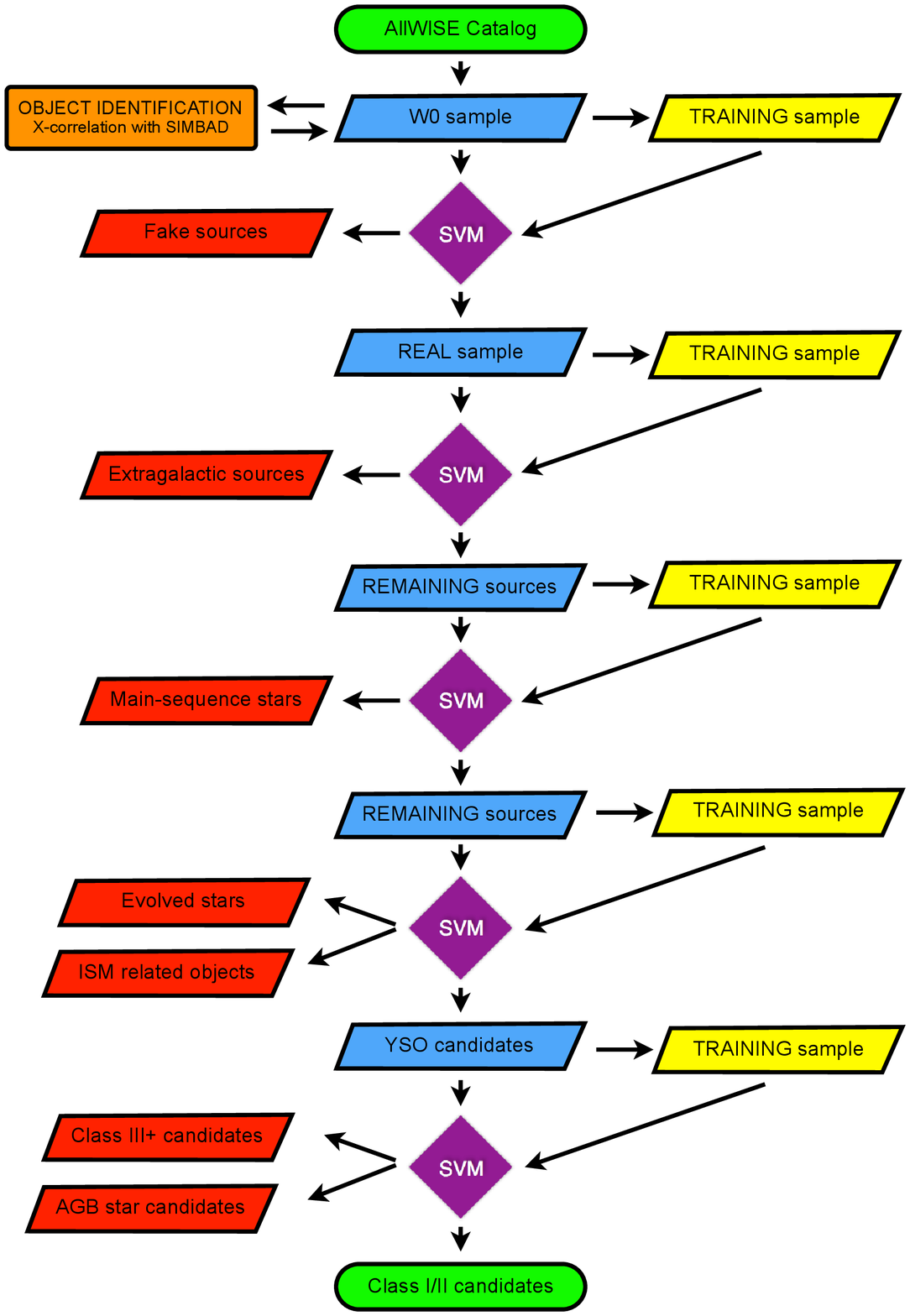}
\caption{Steps from the AllWISE Catalogue to the final Class I/II candidate selections. Steps are described in details in the corresponding subsections.}
\label{dectree}
\end{figure*}

A multi-step classification scheme was developed to remove the contaminating sources from our sample, and to identify the YSO candidates with high accuracy. The steps of our classification are shown in Figure \ref{dectree}. Each step and the corresponding training samples are described in detail below. Although SVM is a very powerful tool, the necessity of a multi-step classification scheme can be explained by considering the complex structure of the ISM. The interstellar reddening has an impact on the apparent colours of the sources, therefore it is important to take into account where different object types are typically found as a function of ISM column density. For this purpose, and by using the Planck dust opacity maps, we binned our sources according to the dust opacity value registered at their position in the sky.

\subsection{Spurious source identification}
\citet{koenig2014} performed a careful examination of the AllWISE catalogue: they inspected, at the positions of the AllWISE sources, the higher resolution Spitzer images. In addition, in selected regions of the sky, they compared Spitzer source catalogues to AllWISE lists of sources. Their analysis led them to conclude that several AllWISE catalogued sources are spurious, and that many are likely ISM knots, or {\em{cirrus}}. Following their finding, our very first step was to identify and remove the spurious sources from our \textit{W0} sample.

To train the SVM we checked the WISE W3 and W4 images of \textit{W0} positions in five different regions: the Galactic centre, the California Molecular Cloud, the Galactic anti-centre, the $\rho$ Oph star forming region and the Cepheus molecular complex. In these regions we selected 680 positions where we were able to clearly identify a point source, and 664 positions where visual source identification was not possible. Examples for these real and spurious AllWISE detections are shown in Figure \ref{realw3w4}  and Figure \ref{fakew3w4}, respectively. The training samples included, for these sources, the AllWISE catalogue information on the signal-to-noise ratio (SNR), the reduced $\chi^2$ value, the number of times the source was detected with SNR$>$3 and the number of profile fits in the W3 and W4 bands. 

\begin{figure}
\includegraphics[width=0.45\textwidth, angle=0]{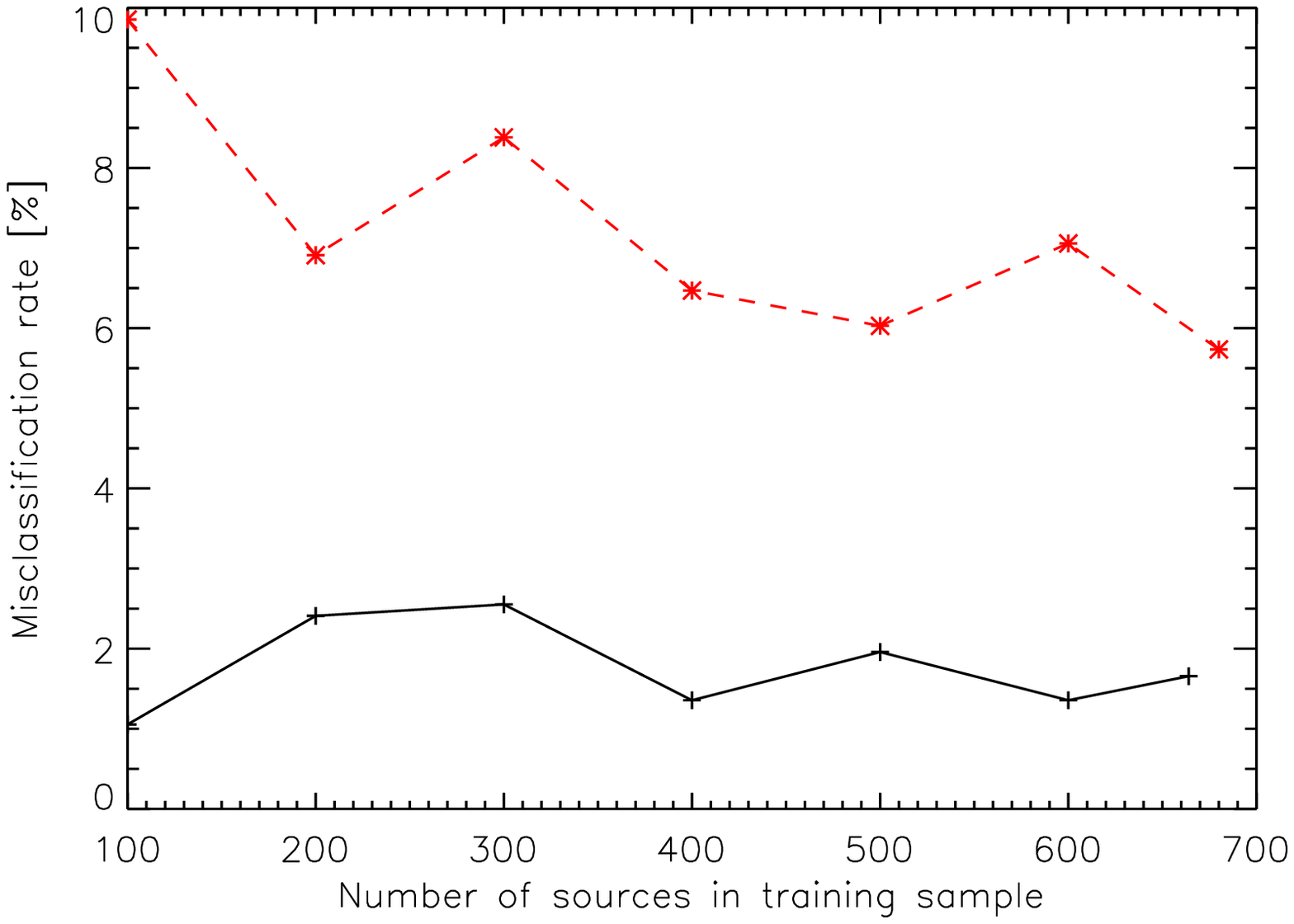}
\caption{The rate of false positive (black solid line) and false negative (red dashed line) as a function of elements used in the training sample. On average only 1.8\% of the spurious sources were classified as a real source.}
\label{spuriousmisclass}
\end{figure}

With the help of the training sample described above, we classified the \textit{W0} sample into two classes: real and spurious sources. The misclassification rate, i.e. the rate of false positive and false negatives, was investigated as a function of the number of elements used in the training sample. Figure \ref{spuriousmisclass} shows that the rate of misclassified sources does not change significantly with the number of elements used. The ratio of false negatives was 7.2$\pm$1.4\% in the test. The lowest misclassification rate was achieved with the maximum number of elements in the training sample (5.7\%). The fraction of false positives was found to be 1.8$\pm$0.6\%. With the maxmimum number of elements in the training sample, it was 1.7\%.
The spurious and real sources were classified in their own class with 98.3\% and 94.3\% success rate, respectively. Applying the determined classification boundaries to our initial sample, we classified 5,366,238 sources as spurious and 3,590,398 sources as real. The latter constitutes what we refer to as \textit{real} sample.

\subsection{YSO identification process}\label{subtypes}

The SIMBAD database lists 235 object types. This large variety makes the training of the algorithm rather inefficient. To make the algorithm more powerful and to have statistically more robust training samples, we binned the object types based on similarities in their J$-$H, H$-$K$_s$, K$_s-$W1, W1$-$W2, W2$-$W3, W3$-$W4 colours and on the \textit{ext} (extended source) parameter. Figures \ref{exgal_ext}--\ref{ism_ext} show the distribution of the average colours and the average \textit{ext} values. 

Three large groups of SIMBAD extragalactic objects were created. SIMBAD types belonging to the G1 group have low H$-$K$_s$ and W1$-$W2 colours and have mostly high \textit{ext} values. Source types of G2 group are less extended based on the \textit{ext} value and have high W1$-$W2 values. The remaining sources were classified as G3, and are mostly extended like the G1 objects but have higher colour indices at shorter wavelengths.

Evolved stars were also grouped in three large bins. E1 type objects have rather small colour indices and are more compact, while E2 types have high J$-$K colours compared to the other object types. E3 objects have W2$-$W3 colour higher than all the other evolved types.

Two bins of young SIMBAD objects were created: All colour indices of Y1 sources are higher than those of Y2 objects. Y1 sources also appear to be less compact because they have higher \textit{ext} values.

The SIMBAD type ``*'' (single star) was not further binned.

Finally, two groups of SIMBAD source types of ISM related objects were defined. ISM1 sources appear to be more compact than ISM2.

The 11 subtypes are listed below, including all SIMBAD types associated to each of them:
\begin{itemize}
\item G1: Galaxy, Part of a Galaxy, Galaxy in Cluster of Galaxies, Brightest Galaxy in a Cluster, Galaxy in Group of Galaxies, Galaxy in Pair of Galaxies, radio Galaxy, HII Galaxy, Low Surface Brightness Galaxy, Emission-line galaxy, Starburst Galaxy, Blue compact Galaxy, LINER-type Active Galaxy Nucleus
\item G2: Active Galaxy Nucleus, Seyfert 2 Galaxy, Blazar, Seyfert Galaxy
\item G3: Broad Absorption Line system, Gravitationally Lensed Image, Gravitationally Lensed Image of a Quasar, Seyfert 1 Galaxy, Quasar, Absorption Line system, Damped Ly-alpha Absorption Line system, Possible Quasar, BL Lac - type object
\item Y1: Young Stellar Object, Variable Star of FU Ori type, Young Stellar Object Candidate
\item Y2: Variable Star of Orion Type, T Tau-type Star, T Tau Star Candidate
\item E1: Horizontal Branch Star, S Star, Red Giant Branch star, Possible Carbon Star, Possible S Star, Carbon Star, Yellow supergiant star, Asymptotic Giant Branch Star, Evolved supergiant star, Variable Star of Mira Cet type, Possible Yellow supergiant star, Possible Horizontal Branch Star, Possible Red supergiant star, Red supergiant star
\item E2: Possible Supergiant star, Possible Asymptotic Giant Branch Star, OH/IR star
\item E3: Post-AGB Star, Post-AGB Star Candidate
\item S: Single star
\item ISM1: Interstellar matter, HI shell, High-velocity Cloud, Emission Object, Planetary Nebula, Possible Planetary Nebula, SuperNova Remnant, SuperNova Remnant Candidate, Dark Cloud (nebula), Cloud, Part of Cloud, Outflow
\item ISM2: Molecular Cloud, HII (ionized) region, Galactic Nebula, Globule (low-mass dark cloud)
\end{itemize}

\subsubsection{Removal of extragalactic sources}
\begin{figure}
 \includegraphics[width=0.45\textwidth,trim=0cm 0cm 0cm 4cm]{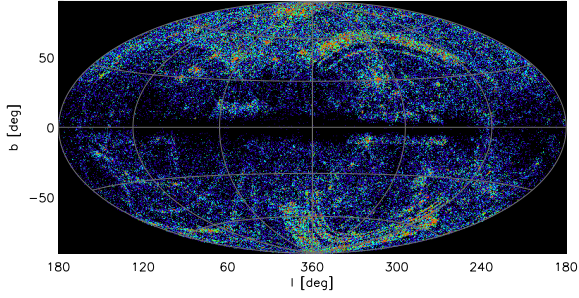}
\caption{Surface density of extragalactic sources identified with SIMBAD, shown in galactic equal-area Aitoff projection. The direction of the Galactic mid-plane is almost completely galaxy-free. Values were calculated in $0.5^{\circ} \times 0.5^{\circ}$ bins, and are represented on linear scale from $0$ to $4$.}
\label{simbadgal}
\end{figure}

\begin{figure}
\includegraphics[trim= 4.0cm 2cm 2.5cm 2cm, angle=270,clip=true,width=0.45\textwidth]{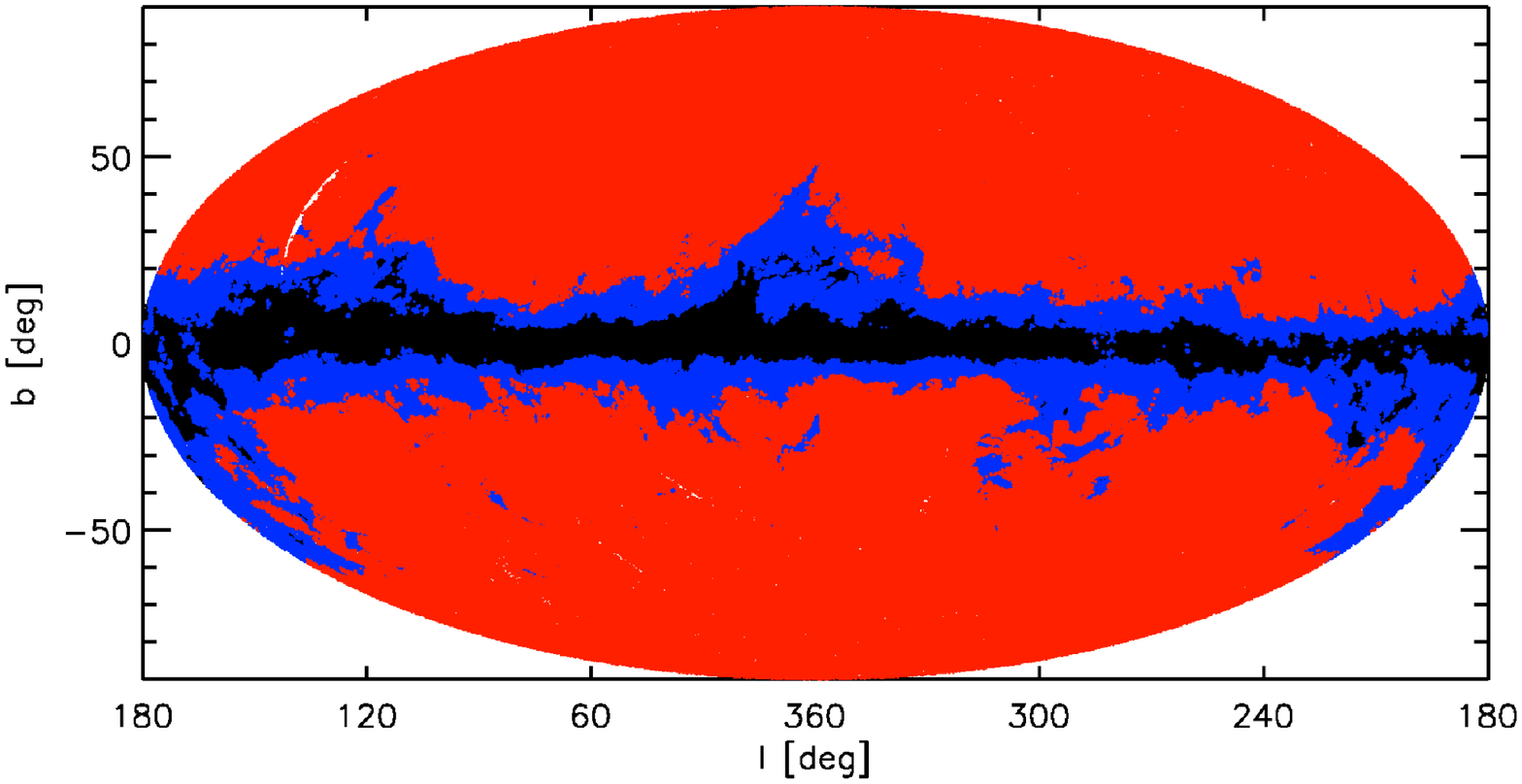}
\caption{Regions of the sky containing 90\% (red), 9\% (blue) and 1\% (black) of the SIMBAD identified extragalactic sources. Regions are represented in galactic equal-area Aitoff projection.}
\label{galebv}
\end{figure}

First, the \textit{real} sample was analysed with the goal of removing the extragalactic sources. We were able to identify 105,564 known extragalactic sources, including all SIMBAD object types that belong to the Galaxy type\footnote{http://cds.u-strasbg.fr/cgi-bin/Otype?X}. As seen in Figure \ref{simbadgal} their surface density distribution in the sky is very inhomogenous. Regions close to the galactic mid-plane and around known giant molecular clouds (GMCs), like Orion (at l$\simeq$200, b$\simeq-$10) or Cepheus (at l$\simeq$110, b$\simeq$10) are almost free of extragalactic sources. These regions contain high amount of ISM compared to the surroundings, therefore the extragalactic source distribution is biased in these regions. On one hand, the ISM is opaque at visual and near-IR wavelengths. Most of the SIMBAD extragalactic counts were made in the visual and near-IR regime, resulting in incomplete catalogues. On the other hand, IR-bright ISM features are able to cover fainter extragalaxies, making them undetectable. Third, the interstellar reddening modifies the apparent colours of the background objects. We found that 90\% of the extragalactic sources are located in regions where $\tau<1.25\times10^{-5}$ and only 1\% are located in regions where $\tau>5.05\times10^{-5}$ (see Figure \ref{galebv}). Nine percent are located in regions between the two values. After investigating which colours and brightness values provide the highest separation between the extragalactic sources and all the other object types, we found that our trainer would provide the best separation if the J$-$H, J$-$W4, K$_s-$W4, W2$-$W3 colours, the J and W2 band magnitudes and the \textit{ext} parameter are used. In each of these three regions (with characteristic $\tau$ values), we performed an SVM classifications using all subtypes listed in the previous section. Those sources classified as either G1, G2 or G3, were identified as extragalactic objects. 

As a result 377,126 sources have been classified as extragalactic sources (G1, G2 or G3), and were removed to build the \textit{galaxy-free} sample. Out of the 105,564 already known extragalactic objects 105,376 were re-classified as extragalactic, and only 188 were misclassified (false-negative). The \textit{real} sample contained 5,685 known YSOs, of which only 117 were classified as extragalactic source (false-positive).

\subsubsection{Removal of main sequence stars}
After removing the sources classified as extragalactic objects our goal was to remove the main-sequence (MS) stars from the remaining 3,213,272 sources. This \textit{galaxy-free} sample contained 932,531 known MS stars. As it was done in case of the galaxy removal, we checked again the $\tau$ distribution of the MS stars and divided the sky into three regions. Those regions contained one third of the known MS stars, having $\tau<1.1\times10^{-4}$,  $\tau>2.8\times10^{-4}$ and in-between. Also (as in the previous step), the average colours and brightness values of the different subtypes were calculated to achieve a maximum possible separation between MS stars and all the other object types. Performing this task led us to use the colours H$-$W4, K$_s-$W4, W1$-$W2, W2$-$W3, W3$-$W4, the H and W4 band magnitude values and the \textit{ext} parameter.

Our results showed that we successfully removed 90.1\% of the known MS stars of the \textit{galaxy-free} sample, by classifying 2,052,410 sources as MS star. At the same time only 327 of the known YSOs were classified as MS star, meaning that 92.5\% of the known YSOs of the \textit{real} sample were still kept.

\subsubsection{Classification into evolved stars, ISM related objects \& YSO candidates}

In the next step we wanted to divide the remaining (\textit{galaxy- and MS star-free}) 1,160,862 sources into three main categories: evolved stars (E1, E2 and E3 subtypes), ISM related objects (ISM1 and ISM2 subtypes) and YSO candidates (Y1 and Y2 subtypes), with the goal to keep sources classified as Y1 or Y2. We were not able to identify regions, where one of the object types was dominant, thus cuts in the $\tau$ value were not applied. The training sample was prepared by using J$-$H, W2$-$W3, W3$-$W4 colours, the W2 magnitude and the \textit{ext} parameter. 

This classification step resulted in losing only 210 of the known YSOs, while keeping 88.8\% of the known YSOs in the \textit{real} sample. We also removed 11,589 of the known evolved stars. Compared to the number of known evolved stars in the \textit{real} sample, 27.4\% of them were still present at this stage. We were also able to remove 62.3\% of the known ISM related objects. The resulting YSO candidate sample (sources classified as Y1 or Y2) contained 751,628 sources.

\subsubsection{Classification into YSO evolutionary classes}\label{classtrain}
The last step was to separate the Class I, II and III sources \citep{lada1987} with the ultimate goal of finding reliable Class I and Class II YSO candidates. Class I and II sources have a significant IR excess that originates from the dust in their circumstellar envelopes or protoplanetary disks. The Class III sources have infrared colours more similar to main sequence stars, but still showing infrared excess. The SIMBAD database does not list information on the actual evolutionary stage of the known YSOs, therefore our training sample was prepared based on YSO catalogs from the literature, preferably listing the evolutionary classes.  We used catalogs from the following papers: \citet{allen2012}, \citet{billot2010}, \citet{chavarria2008}, \citet{connelley2008}, \citet{evans2009}, \citet{gutermuth2008}, \citet{gutermuth2009}, \citet{kirk2009}, \citet{koenig2008}, \citet{megeath2012}, \citet{rebull2011}, \citet{rivera2011}.  We note that these papers and their classification methods are based on IR data, mainly obtained with the Spitzer Space Telescope \citep{werner2004}. We also used \citet{connelley2010} and \citet{fang2009}, which are catalogues of spectroscopically confirmed YSOs, and \citet{winston2007} and \citet{winston2010} listing YSOs with X-ray data.

Based on these catalogues, we created a training sample that contained 247 Class I, 1925 Class II and 313 Class III objects (the latter category includes sources of \citet{koenig2008}, being stars in star forming regions, but showing mostly photospheric colours). In order to simplify the classification, where listed, we considered Transition Disk and Flat SED objects as Class II sources. 

The first attempt to classify our YSO candidates into evolutionary stages failed, and the contamination caused by asymptotic giant branch star candidates (``AB?'' as listed in SIMBAD) was still high. Therefore, the following step was additionally performed: the SVM was trained by using three subtypes, the Class I/II, the ClassIII+ and the ``AB?'' stars. Majority of the ``AB?'' objects are those identified by \citet{robitaille2008}, and we disagree with their statement "YSOs and AGB stars can be mostly separated by simple colour--magnitude selection criteria". The J$-$H, H$-$W3, W1$-$W2, W1$-$W4 and W2$-$W3 colours were used along with the W2 magnitude and the \textit{ext} parameter, to create our training sample.

\begin{figure*}
\includegraphics[width=\textwidth,trim=0cm 0cm 0cm 6cm]{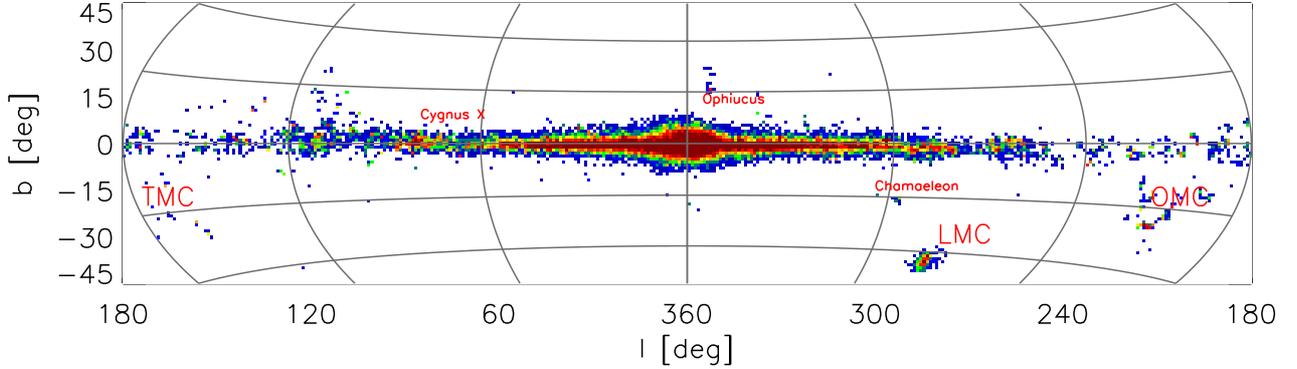}
\caption{Surface density of the Class I/II YSO candidate sources classified with SVM, shown in galactic equal-area Aitoff projection. Values were calculated in $1^{\circ} \times 1^{\circ}$ bins, and are represented on linear scale from $1$ to $100$. The major and best known star forming regions can be easily identified.}
\label{ysocandidatessd}
\end{figure*}

\begin{figure}
\includegraphics[width=0.5\textwidth, angle=0]{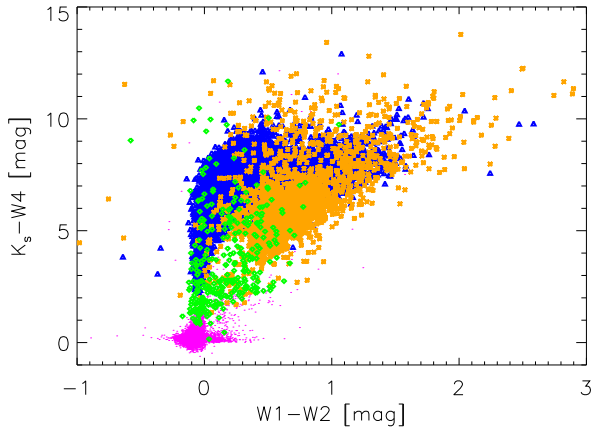}
\caption{W1-W2 vs. K$_s$-W4 colour-colour diagram with our Class I/II candidate sources (orange crosses) overlaid on sources classified as Class III+ (green diamonds), known main sequence stars from SIMBAD (magenta dots) and known SIMBAD extragalactic sources (blue triangles).}
\label{classifiedysos}
\end{figure}

As a final result we classified $133,980$ sources as Class I/II candidates. Their surface density distribution is shown on Figure \ref{ysocandidatessd}. Compared to the training sample of 247 Class I, 1925 Class II and 313 Class III objects, the resulting Class I/II candidate sample contains 240 of the known Class I, 1824 of the known Class II and 79 of the known Class III sources. This means that 95\% of the known Class I+II sources were successfully kept, while 74.8\% of the Class III sources were removed. Likewise, in this last step 63.7\% of the known AGB star candidates were also removed. Figure \ref{classifiedysos} illustrates the robustness of our method, as it shows the significant overlap between the known extragalactic sources, the known field stars, and our samples classified as Class I/II candidates and Class III+ candidates.

\section{Discussion}

\subsection{Reliability, false-positives}\label{falsepositive}
We carefully investigated the possible contamination present in our Class I/II candidate sample. In our \textit{real} sample the number of sources identified with SIMBAD was 1,151,956, including 5,685 YSOs (sources with object types ``Y*O'', ``TT*'', ``Or*'', ``FU*'', ``Y*?'' or ``TT?''). We found only 21,568 (1.9\%) false-positive classifications that remained in our SVM classified Class I/II candidate sample. This means that we were able to remove 98.1\% of the contamination, as compared with the SIMBAD training set. The 21,568 false-positive sources were further analysed in order to learn what fraction of the contamination is coming from the different SIMBAD object types. The complete list is shown in column "SVM" in Table \ref{simbadcomparison}. Here we created four different categories to summarize our findings, representing groups of object types that are (i) most probably contamination, like known extragalactic objects or known evolved stars, (ii) candidate SIMBAD object types, (iii) sources that are assigned a generic object type in SIMBAD, such as IR source or star, but for which there is a non-zero probability to be instead YSOs and (iv) sources of flux that might be YSOs, or are closely related to them (e.g. maser).

\begin{enumerate}
\item Obvious contamination
	\begin{itemize}
	\item Extragalactic source -- 101 (/105,564 -- 0.1\%)
	\item Evolved star -- 830 (/13,121 -- 6.3\%)
	\item Other -- 1,955
	\end{itemize}
\item Possible contamination -- candidate object type -- 1,128 (/7,033 -- 16\%)
	Here we have to note that 931 of the 1128 candidate type objects are from the ``AB?'' type. 925 of them are classified as ``AB?'' by \citet{robitaille2008}.
\item Possible YSOs
	\begin{itemize}
	\item Single star -- 11,990 (/932,531 -- 1.3\%)
	\item Infrared source -- 1,618 (/6,701 -- 24.1\%)
	\item Sources related to ISM -- 413 (/1,226 -- 33.7\%)
	\item Variable star -- 1,424 (/16,537 -- 8.6\%)
	\end{itemize}
\item Possibly related to YSOs
	\begin{itemize}
	\item Radio, mm, sub-mm or X-Ray source -- 185 (/3,089 -- 6\%)
	\item Maser 49 (/71 -- 69 \%)
	\end{itemize}
\end{enumerate}

As it is shown in the list, the total number of obvious and possible contamination is very low in the candidate sample, 4013 (0.35\%) sources in total. The number of obvious contaminants only, is even lower, 2,886 sources (0.25\%).

We also made a comparison based on the binned SIMBAD subtypes defined in Section \ref{subtypes}. Table \ref{subtypemiss} details the number of elements for each subtype, found in the samples indicated by the table columns. We notice that only 101 sources (0.1\%) of the galaxy subtypes (G1, G2 and G3) are still present in the final Class I/II sample. The remaining number of evolved stars was found to be 1792 that is 11.2\% of the total E1+E2+E3 subtypes in the \textit{real} sample. We note again that 925 of the 1792 (52\%) are asymptotic giant branch star candidates (``AB?'') of \citet{robitaille2008}. Without these objects the total number of E1+E2+E3 sources would be only 867. The fraction of remaining single stars (as listed in SIMBAD) is only 1.3\%. This small fraction corresponds to 11,990 sources, which is higher than the number of SIMBAD YSOs, but it is also a very generic object type that does not prevent these sources from being YSOs in reality. The number of objects from ISM related types (ISM1+ISM2) is 356, corresponding to 33.5\% of the \textit{real} sample. We emphasise once again that the SIMBAD associations are rather generic, therefore some of the sources might actually turn out to be YSOs.

\begin{table}
\begin{tabular}{lccc}
subtype&W0&real&Class I/II\\
\hline
G1	&148,267	&90,840	&65		\\
G2	&10,311	&6,152	&7		\\
G3	&12,729	&8,572	&29		\\
E1	&13,208	&12,940	&631		\\
E2	&3,429	&2,925	&1,118	\\
E3	&109		&106		&43		\\
ISM1	&912		&745		&183		\\
ISM2&515		&319		&173		\\
S	&973,629	&932,733&11,990	\\
Y1	&9,268	&4,930	&3,705	\\
Y2	&1,128	&755		&637		\\
\end{tabular}

\caption{Number of known sources in our samples of the selection process. First column indicates the name of the subtype (as defined in Section \ref{subtypes}). Column 2, 3 and 4 are the \textit{W0}, the \textit{real} and the final Class I/II candidate samples.}\label{subtypemiss}
\end{table}

The Sloan Digital Sky Survey DR-9 \citep[SDSS DR-9]{adelman2012} flags the object types for all their sources indicating whether the source is thought to be a galaxy or a star based on morphology. A cross-correlation with SDSS DR-9 allowed us a different estimation on the fraction of false-positive classifications. Of the SIMBAD YSOs in the \textit{real} sample, 1,029 of the 5,685 known YSOs have a counterpart in SDSS DR-9 (using a searching radius of 5$^{\prime\prime}$). 14\% of them (144) were flagged as galaxy in their catalog. We also cross-checked the known YSOs in our final Class I/II candidate sample. 814 of them were found in the SDSS of which 106 (13\%) are flagged as galaxy. Out of the total $133,980$ sources classified as Class I/II, 5,840 were found in the SDSS catalogue, 580 of them are flagged as galaxy (9.9\%). We conclude that our final SVM sample of candidate Class I/II does not contain a higher faction of extragalactic contaminants than the sample of known YSOs in SIMBAD.

\subsection{Completeness, false-negatives}\label{falsenegative}

In the process of identifying the Class I/II sources candidates, a number of known YSOs were lost by either misclassification (false negatives) or because they were not detected by WISE. The completeness of our sample was investigated in three different ways. (i) First, we searched our \textit{W0}, \textit{real} and final Class I/II sample for the known YSOs in SIMBAD; (ii) then we looked, in the same samples as in i), for YSOs listed in public photometric catalogues (see \ref{classtrain}) (iii) finally, again using the samples as in i), we looked for spectroscopically confirmed YSOs.

\subsubsection{Comparison to SIMBAD YSOs}
First, we considered all the known SIMBAD YSOs and searched for them in our \textit{W0}, \textit{real} and final Class I/II samples. The total number of known YSOs in the SIMBAD database was 46,453 at the time of our investigation (21,186 ``Y*O'', 20,716 ``Y*?'', 1,831 ``TT*'', 237 ``TT?'', 33 ``FU*'' and 2,450 ``Or*''). After cross-correlating with SIMBAD, we found a total number of 10,396 known YSOs (4,496, 4,039, 733, 61, 20 and 1,047, respectively) in our \textit{W0} sample - this represents the number of known YSOs observed by WISE and detected above S/N$>$3 in all WISE bands with 2MASS photometric errors below 0.1 mag. After the spurious source identification our \textit{real} sample contained 5,685 of the known YSOs. The number of known YSOs in the final Class I/II sample is 4,342. Based on this comparison, the completeness of our Class I/II sample is 9.3\% compared to the total number of YSOs known to SIMBAD. The completeness is 40.4\% compared to the \textit{W0} sample, and 61.7\% compared to the \textit{real} sample. The distributions for the normalised magnitude of the known YSOs in the \textit{W0}, \textit{real} and Class I/II samples are shown in Table \ref{completenesshist}. These plots reveal that the known YSOs lost in the selection process are mainly located at the faint end of the magnitude distributions, suggesting that our selection method is more sensitive to the brighter YSOs.

\subsubsection{Comparison to photometric YSO catalogues}\label{test2}
The next step consisted in searching in our \textit{W0}, \textit{real} and Class I/II samples for YSOs listed in published photometric catalogues. A separate search in the literature was performed for Class I\&II and Class III sources. The results are listed in Table \ref{photocomp}. Due to the initial selection conditions, only 36.4\% of the literature Class I\&II sources and 30.4\% of the literature Class III sources were found in the \textit{W0} sample. The \textit{real} sample contains 20.8\% of the Class I\&II and 12.0\% of the Class III sources. Our final Class I/II contains 12.9\% of the literature Class I\&II sources and only 3.3\% of the literature Class III sources. In addition, 52.6\% of the Class I\&II sources in the textit{W0} sample are classified again as such, while 10.8\% of the Class III sources are erroneously classified as Class I/II. For the real sample, 92.8\% of the Class I\&II sources were successfully classified, while 27.5\% of the Class III sources were re-classified as Class I/II.

\subsubsection{Comparison to spectroscopic YSO catalogues}
Finally, the steps described in Section  \ref{test2} were repeated in the case of spectroscopically confirmed YSOs, which are mostly located in nearby star forming regions. The results of the comparison are provided in Table \ref{speccomp}. Only about 42\% of the spectroscopically confirmed YSOs fulfilled our initial W0 condition. This also shows the limitations of a WISE--2MASS selection, i.e. a large proportion of YSOs do not have reliable fluxes in all WISE and/or 2MASS bands. An additional 249 sources (14.4\%) turned out to be spurious WISE detections, while 434 of the remaining 475 (91.3\%) were successfully re-classified.

\begin{table*}
\footnotesize\addtolength{\tabcolsep}{-4pt}
\begin{tabular}{||cl|cc|cc|cc|cc|}
&Region&\multicolumn{2}{c|}{Paper}&\multicolumn{2}{c|}{W0}&\multicolumn{2}{c|}{Real}&\multicolumn{2}{c|}{SVM Class I/II}\\
&&Class I/II&Class III&Class I/II&Class III&Class I/II&Class III&Class I/II&Class III\\
\hline
\hline
\citet{allen2012}& Cepheus OB3 &		1135&	1440&	435&		279&		209&		29&		198&	9\\
\citet{billot2010}& Vul OB1 &		703&		153&		259&		82&		160&		54&		134&	7\\
\citet{chavarria2008}&S254-S258&	252&		210&		63&		42&		18&		10&		16&	0\\
\citet{connelley2008}&--&	220&		--&		51&		--&		47&		--&		44&	\\
\citet{evans2009}&--&		942&		79&		294&		37&		214&		28&		204&	4\\
\citet{gutermuth2008}&NGC 1333&	133&		--&		57&		--&		46&		--&		43&	--\\
\citet{gutermuth2009}&--&	2548	&	--&		926&		--&		549&		--&		518&--	\\
\citet{kirk2009}&Cepheus flare&		128&		13&		97&		10&		76&		5&		69&	1\\
\citet{megeath2012}&Orion A\&B&		2284&	329&		1023&	168&		631&		69&		603&	53\\
\citet{rebull2011}&NAN complex&		1149&	112&		498&		98&		264&		84&		217&	1\\
\citet{rivera2011}&W3 molecular cloud&		1566	&	--&		312&		--&		57&		--&		55&--\\
\citet{winston2007}&Serpens cloud core&	115&		22&		46&		2&		31&		2&		22&	1\\
\citet{winston2010}&NGC 1333&	54&		41&		36&		11&		34&		6&		31&	3\\
\hline												
Total&	&	11229&	2399&	4097&	729&		2336&	287&		2154&79\\
\end{tabular}
\caption{Comparison of our selection to existing photometry--based YSO catalogues. Column 1 lists the corresponding paper. Column 2 lists the region of the sky that was the subject of the study. Column 3 and 4 give the number of Class I/II and Class III YSOs listed in the paper. Column 5 and 6 list the number of Class I/II and Class III objects found in the \textit{W0} sample. Columns 7 and 8, and columns 9 and 10 give the number of corresponding objects in the \textit{real} and Class I/II sample. \label{photocomp}}
\end{table*}

\begin{table*}
\begin{tabular}{| l | l | c | c | c | c |}
&Region&Paper&W0&Real&SVM Class I/II\\
\hline
\hline
\citet{alcala2014}&Lupus&		36&		35&		28&		25\\
\citet{an2011}&Galactic Center&			35&		1&		1&		1\\
\citet{connelley2010}&--&	88&		38&		37&		36\\
\citet{cooper2013}&--&		180&		49&		48&		48\\
\citet{erickson2015}&Serpens&		63&		33&		22&		18\\
\citet{fang2009}&L1630N\&L1641&		330&		183&		106&		100\\
\citet{kumar2014}&Carina nebula&		241&		23&		2&		1\\
\citet{kun2009}&Cepheus flare&		77&		68&		68&		63\\
\citet{mooley2013}&Taurus--Auriga&		13&		10&		9&		4\\
\citet{oliveira2009}&Serpens&		58&		49&		46&		40\\
\citet{rebollido2015}&$\rho$ Oph&	48&		48&		44&		39\\
\citet{szegedielek2013} (a)&ONC&	372&		131&		55&		52\\
\citet{szegedielek2013} (b)&ONC&	187&		56&		9&		7\\
\hline												
Total&&		1728&	724&		475&		434\\
\end{tabular}
\caption{Comparison of our selection to existing spectroscopic YSO catalogues. The first column list the papers used in the analysis. Column 2 lists the corresponging sky region. Column 3 gives the number of YSOs used from the paper. The fourth column lists the number of YSOs found in our \textit{W0} sample. Column 5 and 6 list the number of YSOs classified as \textit{real}, and further classified as Class I/II YSO candidate. (a) and (b) stands for the Classical T Tauri and weak-line T Tauri objects, respectively. \label{speccomp}}
\end{table*}

\subsection{Comparison with the \citet{koenig2014} method}

Recently \citet[hereafter KL14]{koenig2014} published a method to reduce the number of spurious sources and to identify potential YSO candidates in the WISE and AllWISE catalog. In this comparison we applied their method to our \textit{W0} sample to be able to compare the results, including the spurious source removal. As a result, the KL14 method identified 124,608 YSO candidates from the \textit{W0} sample.

In Table \ref{subtypemisskonig} we compare the number of sources of our 11 binned subtypes in the following three samples: (1) the initial \textit{W0} sample (S/N$>$3 in all WISE bands and 2MASS photometric errors < 0.1), (2) our SVM selected Class I/II sample and (3) the YSO sample resulted from applying the KL14 method to the initial \textit{W0} sample. Because the KL14 method has its own criteria for spurious source mitigation, the comparison is fair if we apply it to our \textit{W0} sample. We can see that the number of false-positive extragalactic objects in our Class I/II selection is 101 (0.05\% compared to the \textit{real} sample), while it is 5,889 (3.4\%) in the KL14 sample. The fraction of known SIMBAD YSOs that were recovered with our method is higher (41.7\%) than with the KL14 one (32.1\%). On the other hand, the KL14 method successfully eliminated 95.9\% of the evolved stars (89.3\% with the SVM). We conclude that KL14 method is very efficient in identifying and removing the galactic contamination, but allows us to retrieve a lower number of known YSOs. Our method is more successful in removing extragalactic contamination, and it misclassifies known YSOs in a smaller fraction. A summary of the comparison between the two methods is provided in Table \ref{simbadcomparison}.

\begin{table}
\begin{tabular}{lccc}
Subtype&W0&SVM Class I/II&KL14 YSOs\\
\hline
G1	&148,267	&65		&1398	\\
G2	&10,311	&7		&1188	\\
G3	&12,729	&29		&3303	\\
E1	&13,208	&631		&58		\\
E2	&3,429	&1,118	&613		\\
E3	&109		&43		&16		\\
ISM1	&912		&183		&90		\\
ISM2&515		&173		&50		\\
S	&973,629	&11,990	&1043	\\
Y1	&9,268	&3,705	&2695	\\
Y2	&1,128	&637		&650		\\
\end{tabular}

\caption{Number of sources of our binned SIMBAD subtypes (first column) in the initial \textit{W0} sample (second column), in our SVM classified Class I/II candidate sample (third column) and in the YSO sample of the KL14 method (last column). KL14 successfully removes the galactic contamination, but is less successful in identifying the known YSOs and in removing the extragalactic contamination.}\label{subtypemisskonig}
\end{table}

We also investigated whether the KL14 method is more sensitive to Class I/II YSOs rather than to more evolved class types. Using the same sources that we used in Section \ref{classtrain} we found that KL14 method finds 95 of the 247 Class I sources (SVM finds 240), 922 of the 1925 Class II sources (1824 with SVM) and 47 of the 313 Class III objects (SVM result is 79). The combined results show that SVM is able to retrieve 2064 of the 2172 Class I/II objects (95\%), while the KL14 method finds only 589 (27\%). These results suggest that the overlap between the two methods is rather small, however the results based on the SIMBAD comparison show a somewhat better agreement. To find out the reason for such a discrepancy, we calculated the average magnitudes in each 2MASS and WISE band for the SIMBAD YSOs found in each selection. The results are listed in Table \ref{simbadysomagnitudes}. We also analysed the magnitude distribution of the SIMBAD YSOs in the KL14 selection (see Figure \ref{completenesshist}). Our conclusion is that the KL14 method is overall more sensitive to the fainter sources than our method. This can explain why, on one side, the KL14 method classifies more extragalactic sources as YSOs and successfully removes bright galactic contaminants while, on the other, it recovers a smaller fraction of YSOs.

\begin{table}
\begin{tabular}{lcc}
Band&KL14 YSOs [mag]&SVM Class I/II [mag]\\
\hline
J		&13.9$\pm$1.7	&13.1$\pm$2.0\\
H		&12.5$\pm$1.4	&11.6$\pm$1.8\\
K$_s$	&11.6$\pm$1.4	&10.7$\pm$1.7\\
W1		&10.6$\pm$1.4	&9.6$\pm$1.7\\
W2		&9.8$\pm$1.5	&8.9$\pm$1.7\\
W3		&7.5$\pm$1.7	&6.6$\pm$1.8\\
W4		&5.0$\pm$1.7	&4.4$\pm$2.0\\
\end{tabular}

\caption{Average brightness of the SIMBAD YSOs classified as YSO with the KL14 method (second column) and with our SVM method (third column). The KL14 method appears to be more sensitive to the fainter sources.}\label{simbadysomagnitudes}
\end{table}

\subsection{Comparison with the Quadratic Discriminant Analysis}

In our previous work \citep{toth2014} the Quadratic Discriminant Analysis \citep[QDA]{mclachlan1992} technique was used to identify YSO candidates using far--IR AKARI and mid--IR WISE data. In this section we compare the currently used SVM method and the QDA by repeating the steps of the analysis described above, and using the same training samples, to find out which one suits the problem better. The main difference between the two methods is their approach to the decision boundaries. Discriminant Analysis techniques perform dimensionality reduction, and project the data into a subspace where they maximise the separation. On the contrary, the SVM maps the data into a higher--dimensional space and a hyperplane is calculated that provides the best separation of the classes.

As a first step the classification of real and spurious sources was repeated. QDA successfully classified as such only 77.9\% of the spurious sources (compared to 98.3\% for SVM), and only 37.9\% of the real sources (compared to 94.2\% for SVM). In this case, SVM clearly outperforms the QDA method.

As a second step we repeated the removal of the extragalactic sources. 99.3\% of the sources in the training sample were successfully re-classified as extragalactic source and only 0.7\% (725) were misclassified by QDA. This is three times more than with SVM, for which only 237 sources were misclassified. By using SVM we were not able to recover 2.1\% of the known SIMBAD YSOs, while we lost 2.3\% of them with QDA.

In the third step we repeated the removal of field stars. With QDA 17.3\% of the SIMBAD single stars remained in our sample, while with SVM only 8.9\%. Also, with SVM 5.5\% of the YSOs were lost while applying QDA to the same training sample resulted in the loss of 7.8\% of them.

As a last step of the QDA and SVM comparison we repeated the classification of the remaining sources into three main object types, YSO candidates, evolved stars and ISM related objects. By using SVM we successfully re-classified 96\% of the known YSOs as YSO candidate. Using QDA the success rate was only 89\%. The contamination caused by the remaining evolved stars is also higher with QDA. The number of evolved objects was found to be 6,194 while it was 4,382 with SVM.

This comparison clearly shows that while QDA and SVM are comparable in some of the steps, the overall performance of SVM is better than QDA, and it is more efficient for classifying Class I/II YSOs.

\subsection{Correlation with the PGCCs}

The Planck Catalog of Galactic Cold Clumps \citep[PGCC]{planck2015} is an all-sky catalogue of Galactic cold clump candidates detected by Planck. The PGCC catalogue contains 13188 Galactic sources spread across the whole sky with a median temperature between 13 and 14.5 K and their size is described with the major and minor full-width half-maximum (FWHM) of a fitted elliptical Gaussian. Cold clumps represent the early stages of star formation  \citep{mckee2007} and a spatial correlation between the cold clump and the YSO distribution is therefore expected.

\begin{figure}
\includegraphics[width=0.45\textwidth, angle=0]{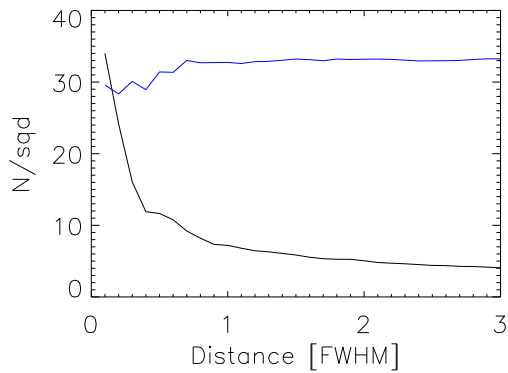}
\caption{Surface density of the Class I/II candidates (black solid line) and the main sequence star candidates (blue line) as a function of the distance relative to the PGCC objects.}
\label{pgcc}
\end{figure}

To further test the robustness of our YSO classification, we analysed their position relative to the PGCCs in the Taurus-Auriga-Perseus-California molecular complex ($150<l<180, -25<b<-1$), which is a well-known star forming region. As a function of the major FWHM we calculated the surface density of our Class I/II candidates and of the objects that we classified as main sequence stars. As seen on Figure \ref{pgcc} the surface density of the Class I/II candidates is highest close to the PGCCs and then rapidly decreases, while that of the main sequence star candidates is independent of the distance measured from the cold clumps. This result strongly suggests that our candidate Class I/II sources are indeed related to the Planck cold clump population.

\section{Summary}

The AllWISE catalog was investigated to identify potential YSO candidates. A subset of the catalogue was used with S/N$>$3 and available 2MASS J, H, K$_s$ data with photometric errors $<$0.1. We applied the Support Vector Machine method to the initial dataset of 8,956,636 sources in a multi--step process. Different combinations of colours and magnitudes were used, in combination with the extended source flag, in order to generate a multi-dimensional training samples and to remove contaminating sources. Sources of known Galactic and extragalactic types were identified with the help of the SIMBAD database, using a 5\arcsec\, radius to match the AllWISE sources. As many as 133,980 objects were classified as YSO Class I/II candidates. The contamination of sources with well-known object types is $<$1\%, in comparison with our SIMBAD training set. We also compared our method to that described in \citet{koenig2014} and to the results obtained with a different approach, the Quadratic Discriminant Analysis. We found that SVM outperforms the \citet{koenig2014} method in preserving the known YSOs and in identifying the extragalactic contamination and it is more effective than QDA.

A positional correlation analysis with the PGCC sources was performed in the case of the Taurus--Auriga--Perseus--California regions. Our Class I/II candidates appear to be characterized by a higher surface density in the proximity of the cold clumps while the main sequence star candidates had a uniform surface density in the field.

\section*{Acknowledgments}
We thank our anonymous referee for all the useful comments that helped us to improve the manuscript. This publication makes use of data products from the Wide-field Infrared Survey Explorer, which is a joint project of the University of California, Los Angeles, and the Jet Propulsion Laboratory/California Institute of Technology, funded by the National Aeronautics and Space Administration. This research has made use of the SIMBAD database, operated at CDS, Strasbourg, France. This research has made use of the VizieR catalogue access tool, CDS, Strasbourg, France. This research was supported by the OTKA grants NN 111016 and K101393.

\appendix
\newpage
\section{Number of SIMBAD object types in the \textit{W0}, Class I/II and KL14 samples.}
\begin{table*}
\footnotesize\addtolength{\tabcolsep}{-4pt}
\begin{tabular}{cccc|cccc|cccc}
SIMBAD	&	\textit{W0}	&	Class I/II	&	Koenig	&	SIMBAD	&	\textit{W0}	&	Class I/II	&	Koenig	&	SIMBAD	&	\textit{W0}	&	Class I/II	&	Koenig	\\
\hline
\hline
unknown	&	7668196	&	113506	&	40034	&	DQ*	&	1	&	0	&	0	&	Or*	&	1048	&	577	&	430\\ 
Rad	&	3440	&	83	&	203	&	AM*	&	3	&	0	&	0	&	RI*	&	136	&	47	&	23\\ 
cm	&	144	&	1	&	21	&	NL*	&	3	&	1	&	0	&	Er*	&	6	&	0	&	0\\ 
mm	&	82	&	27	&	4	&	No*	&	33	&	4	&	4	&	Fl*	&	302	&	40	&	31\\ 
smm	&	40	&	14	&	5	&	DN*	&	18	&	5	&	4	&	FU*	&	20	&	19	&	9\\ 
HI	&	33	&	0	&	0	&	XB*	&	4	&	0	&	0	&	RC*	&	94	&	27	&	21\\ 
Mas	&	79	&	49	&	2	&	LXB	&	4	&	1	&	0	&	Ro*	&	1749	&	5	&	0\\
\end{tabular}
\caption{Comparison of SIMBAD sources found in our initial \textit{W0} sample, in the SVM selected Class I/II YSO sample and in the YSO candidate sample identified by the \textbf{KL14} method. The resolved acronyms of object types can be found here: http://cdsweb.u-strasbg.fr/cgi-bin/Otype?IR \label{simbadcomparison}}
\end{table*}

\newpage
\section{Average colours and magnitudes of SIMBAD object types}
\begin{table*}
\footnotesize\addtolength{\tabcolsep}{-4pt}
\begin{tabular}{cllllllllllllllllllllllllllll}

SIMBAD&J-H&J-K$_s$&J-W1&J-W2&J-W3&J-W4&H-K$_s$&H-W1&H-W2&H-W3&H-W4\\
\hline
Rad& 0.83$\pm$ 0.33& 1.46$\pm$ 0.49& 2.32$\pm$ 0.72& 2.89$\pm$ 0.94& 6.25$\pm$ 1.33& 8.67$\pm$ 1.54& 0.63$\pm$ 0.26& 1.50$\pm$ 0.56& 2.06$\pm$ 0.83& 5.43$\pm$ 1.29& 7.84$\pm$ 1.50\\
 cm& 0.75$\pm$ 0.22& 1.42$\pm$ 0.38& 2.50$\pm$ 0.67& 3.33$\pm$ 0.85& 5.97$\pm$ 1.16& 8.30$\pm$ 1.24& 0.67$\pm$ 0.28& 1.75$\pm$ 0.58& 2.59$\pm$ 0.77& 5.22$\pm$ 1.11& 7.55$\pm$ 1.20\\
 mm& 1.82$\pm$ 0.85& 2.71$\pm$ 1.32& 3.83$\pm$ 1.62& 4.30$\pm$ 1.92& 7.10$\pm$ 2.69&10.20$\pm$ 3.09& 0.89$\pm$ 0.51& 2.01$\pm$ 1.05& 2.48$\pm$ 1.48& 5.28$\pm$ 2.61& 8.38$\pm$ 3.10\\
smm& 1.76$\pm$ 1.10& 2.76$\pm$ 1.65& 3.97$\pm$ 1.94& 4.63$\pm$ 2.27& 7.56$\pm$ 2.19&10.57$\pm$ 2.53& 1.00$\pm$ 0.60& 2.21$\pm$ 1.17& 2.87$\pm$ 1.60& 5.80$\pm$ 2.13& 8.81$\pm$ 2.43\\
 HI& 0.64$\pm$ 0.18& 1.03$\pm$ 0.24& 1.85$\pm$ 0.36& 2.00$\pm$ 0.48& 5.61$\pm$ 1.11& 7.83$\pm$ 1.41& 0.39$\pm$ 0.20& 1.21$\pm$ 0.38& 1.36$\pm$ 0.49& 4.97$\pm$ 1.12& 7.19$\pm$ 1.42\\
Mas& 2.25$\pm$ 1.17& 3.66$\pm$ 1.98& 5.30$\pm$ 2.53& 6.41$\pm$ 3.04& 9.25$\pm$ 2.82&11.71$\pm$ 2.68& 1.41$\pm$ 0.84& 3.05$\pm$ 1.57& 4.17$\pm$ 2.14& 7.01$\pm$ 2.24& 9.47$\pm$ 2.37\\
 IR& 1.31$\pm$ 0.72& 2.04$\pm$ 1.15& 2.84$\pm$ 1.58& 3.20$\pm$ 1.92& 5.32$\pm$ 2.59& 7.11$\pm$ 3.21& 0.74$\pm$ 0.47& 1.53$\pm$ 1.02& 1.90$\pm$ 1.40& 4.01$\pm$ 2.37& 5.80$\pm$ 3.10\\
\end{tabular}

\caption{Average colours and magnitudes of SIMBAD object types} 
\label{simbadcolours1}

\end{table*}

\begin{table*}
\footnotesize\addtolength{\tabcolsep}{-4pt}
\begin{tabular}{cllllllllllllllllllllllllllll}

SIMBAD&J-H&J-K$_s$&J-W1&J-W2&J-W3&J-W4&H-K$_s$&H-W1&H-W2&H-W3&H-W4\\
\hline
LXB& 0.82$\pm$ 0.28& 1.12$\pm$ 0.32& 2.82$\pm$ 1.78& 2.73$\pm$ 2.49& 4.91$\pm$ 3.20& 6.67$\pm$ 3.84& 0.30$\pm$ 0.14& 2.00$\pm$ 2.00& 1.92$\pm$ 2.69& 4.09$\pm$ 3.35& 5.86$\pm$ 3.98\\
HXB& 0.60$\pm$ 0.63& 1.02$\pm$ 0.96& 1.68$\pm$ 1.32& 1.97$\pm$ 1.47& 3.31$\pm$ 2.58& 4.89$\pm$ 3.46& 0.42$\pm$ 0.36& 1.08$\pm$ 0.80& 1.37$\pm$ 0.95& 2.71$\pm$ 2.23& 4.30$\pm$ 3.13\\
ISM& 0.76$\pm$ 0.73& 1.17$\pm$ 1.17& 1.95$\pm$ 1.60& 2.19$\pm$ 2.00& 5.05$\pm$ 2.73& 7.91$\pm$ 3.07& 0.41$\pm$ 0.47& 1.19$\pm$ 1.03& 1.43$\pm$ 1.46& 4.29$\pm$ 2.43& 7.15$\pm$ 2.85\\
PoC& 1.26$\pm$ 0.97& 1.94$\pm$ 1.49& 2.80$\pm$ 1.82& 3.17$\pm$ 2.08& 6.29$\pm$ 2.84& 8.75$\pm$ 2.99& 0.68$\pm$ 0.54& 1.54$\pm$ 1.02& 1.91$\pm$ 1.38& 5.03$\pm$ 2.56& 7.49$\pm$ 2.95\\
PN?& 0.96$\pm$ 0.52& 1.73$\pm$ 0.90& 2.86$\pm$ 1.46& 3.49$\pm$ 1.89& 6.91$\pm$ 2.28& 9.89$\pm$ 2.64& 0.77$\pm$ 0.47& 1.89$\pm$ 1.13& 2.53$\pm$ 1.59& 5.95$\pm$ 2.19& 8.93$\pm$ 2.67\\
bub& 1.61$\pm$ 0.77& 2.39$\pm$ 1.13& 3.38$\pm$ 1.37& 3.76$\pm$ 1.55& 7.06$\pm$ 2.25&10.70$\pm$ 2.67& 0.78$\pm$ 0.38& 1.77$\pm$ 0.85& 2.15$\pm$ 1.11& 5.44$\pm$ 2.13& 9.09$\pm$ 2.58\\
EmO& 0.74$\pm$ 0.55& 1.29$\pm$ 0.89& 2.28$\pm$ 1.23& 2.73$\pm$ 1.50& 5.70$\pm$ 2.32& 8.21$\pm$ 2.83& 0.55$\pm$ 0.41& 1.54$\pm$ 0.92& 1.99$\pm$ 1.21& 4.96$\pm$ 2.23& 7.47$\pm$ 2.79\\
\end{tabular}

\caption{Average colours and magnitudes of SIMBAD object types} 
\label{simbadcolours2}

\end{table*}

\begin{table*}
\footnotesize\addtolength{\tabcolsep}{-4pt}
\begin{tabular}{cllllllllllllllllllllllllllll}

SIMBAD&J-H&J-K$_s$&J-W1&J-W2&J-W3&J-W4&H-K$_s$&H-W1&H-W2&H-W3&H-W4\\
\hline
RC*& 0.96$\pm$ 0.64& 1.85$\pm$ 1.15& 3.29$\pm$ 1.84& 4.33$\pm$ 2.09& 5.96$\pm$ 2.24& 6.75$\pm$ 2.35& 0.89$\pm$ 0.54& 2.33$\pm$ 1.31& 3.36$\pm$ 1.61& 4.99$\pm$ 1.83& 5.79$\pm$ 1.96\\
Ro*& 0.54$\pm$ 0.21& 0.70$\pm$ 0.28& 0.81$\pm$ 0.32& 0.79$\pm$ 0.32& 0.92$\pm$ 0.46& 1.18$\pm$ 0.79& 0.16$\pm$ 0.07& 0.27$\pm$ 0.14& 0.25$\pm$ 0.17& 0.38$\pm$ 0.36& 0.64$\pm$ 0.75\\
a2*&-0.02$\pm$ 0.10&-0.01$\pm$ 0.12& 0.02$\pm$ 0.17& 0.02$\pm$ 0.17&-0.05$\pm$ 0.21& 0.15$\pm$ 0.49& 0.02$\pm$ 0.04& 0.04$\pm$ 0.11& 0.04$\pm$ 0.11&-0.02$\pm$ 0.16& 0.18$\pm$ 0.46\\
El*& 0.09$\pm$ 0.20& 0.15$\pm$ 0.26& 0.20$\pm$ 0.33& 0.23$\pm$ 0.34& 0.27$\pm$ 0.57& 0.76$\pm$ 1.37& 0.05$\pm$ 0.07& 0.10$\pm$ 0.18& 0.14$\pm$ 0.19& 0.17$\pm$ 0.45& 0.67$\pm$ 1.29\\
Psr& 0.82$\pm$ 0.58& 1.14$\pm$ 0.84& 1.65$\pm$ 1.12& 1.66$\pm$ 1.31& 2.81$\pm$ 1.76& 4.39$\pm$ 2.33& 0.32$\pm$ 0.28& 0.83$\pm$ 0.72& 0.84$\pm$ 0.98& 1.99$\pm$ 1.42& 3.57$\pm$ 2.08\\
BY*& 0.47$\pm$ 0.15& 0.61$\pm$ 0.20& 0.69$\pm$ 0.27& 0.73$\pm$ 0.29& 0.78$\pm$ 0.57& 1.01$\pm$ 1.03& 0.14$\pm$ 0.06& 0.22$\pm$ 0.16& 0.26$\pm$ 0.18& 0.31$\pm$ 0.49& 0.54$\pm$ 0.96\\
RS*& 0.50$\pm$ 0.12& 0.64$\pm$ 0.16& 0.74$\pm$ 0.26& 0.79$\pm$ 0.28& 0.82$\pm$ 0.43& 0.96$\pm$ 0.60& 0.15$\pm$ 0.06& 0.24$\pm$ 0.19& 0.29$\pm$ 0.21& 0.32$\pm$ 0.37& 0.46$\pm$ 0.55\\
\end{tabular}

\caption{Average colours and magnitudes of SIMBAD object types} 
\label{simbadcolours3}

\end{table*}

\begin{table*}
\footnotesize\addtolength{\tabcolsep}{-4pt}
\begin{tabular}{cllllllllllllllllllllllllllll}

SIMBAD&K$_s$-W1&K$_s$-W2&K$_s$-W3&K$_s$-W4&W1-W2&W1-W3&W1-W4&W2-W3&W2-W4&W3-W4\\
\hline
Rad& 0.87$\pm$ 0.44& 1.43$\pm$ 0.70& 4.80$\pm$ 1.20& 7.21$\pm$ 1.42& 0.56$\pm$ 0.38& 3.93$\pm$ 1.03& 6.35$\pm$ 1.27& 3.36$\pm$ 0.92& 5.78$\pm$ 1.15& 2.42$\pm$ 0.61\\
 cm& 1.08$\pm$ 0.45& 1.91$\pm$ 0.65& 4.55$\pm$ 1.02& 6.88$\pm$ 1.12& 0.83$\pm$ 0.33& 3.47$\pm$ 0.84& 5.80$\pm$ 0.99& 2.64$\pm$ 0.63& 4.97$\pm$ 0.82& 2.33$\pm$ 0.34\\
 mm& 1.12$\pm$ 0.79& 1.60$\pm$ 1.24& 4.39$\pm$ 2.57& 7.49$\pm$ 3.09& 0.48$\pm$ 0.61& 3.27$\pm$ 2.10& 6.37$\pm$ 2.74& 2.80$\pm$ 1.84& 5.89$\pm$ 2.49& 3.10$\pm$ 1.26\\
smm& 1.21$\pm$ 0.83& 1.86$\pm$ 1.26& 4.80$\pm$ 2.14& 7.81$\pm$ 2.41& 0.66$\pm$ 0.61& 3.59$\pm$ 1.59& 6.61$\pm$ 1.96& 2.93$\pm$ 1.56& 5.95$\pm$ 1.82& 3.01$\pm$ 1.19\\
 HI& 0.82$\pm$ 0.34& 0.97$\pm$ 0.44& 4.58$\pm$ 1.09& 6.80$\pm$ 1.41& 0.15$\pm$ 0.14& 3.76$\pm$ 0.90& 5.98$\pm$ 1.24& 3.61$\pm$ 0.83& 5.83$\pm$ 1.18& 2.22$\pm$ 0.52\\
Mas& 1.64$\pm$ 0.97& 2.75$\pm$ 1.53& 5.60$\pm$ 2.00& 8.05$\pm$ 2.39& 1.12$\pm$ 0.84& 3.96$\pm$ 1.48& 6.42$\pm$ 2.11& 2.84$\pm$ 1.40& 5.30$\pm$ 2.19& 2.46$\pm$ 1.12\\
 IR& 0.80$\pm$ 0.67& 1.16$\pm$ 1.03& 3.27$\pm$ 2.20& 5.06$\pm$ 2.98& 0.36$\pm$ 0.49& 2.48$\pm$ 1.80& 4.27$\pm$ 2.66& 2.12$\pm$ 1.62& 3.91$\pm$ 2.51& 1.79$\pm$ 1.13\\
\end{tabular}

\caption{Average colours and magnitudes of SIMBAD object types} 
\label{simbadcolours4}

\end{table*}

\begin{table*}
\footnotesize\addtolength{\tabcolsep}{-4pt}
\begin{tabular}{cllllllllll}

SIMBAD&K$_s$-W1&K$_s$-W2&K$_s$-W3&K$_s$-W4&W1-W2&W1-W3&W1-W4&W2-W3&W2-W4&W3-W4\\
\hline

LXB& 1.70$\pm$ 1.94& 1.62$\pm$ 2.61& 3.79$\pm$ 3.29& 5.55$\pm$ 3.96&-0.09$\pm$ 0.78& 2.08$\pm$ 1.66& 3.85$\pm$ 2.58& 2.17$\pm$ 1.42& 3.94$\pm$ 2.55& 1.77$\pm$ 1.24\\
HXB& 0.66$\pm$ 0.58& 0.95$\pm$ 0.71& 2.29$\pm$ 2.05& 3.87$\pm$ 2.97& 0.29$\pm$ 0.24& 1.63$\pm$ 1.59& 3.21$\pm$ 2.57& 1.34$\pm$ 1.45& 2.93$\pm$ 2.45& 1.58$\pm$ 1.23\\
ISM& 0.78$\pm$ 0.70& 1.03$\pm$ 1.11& 3.88$\pm$ 2.26& 6.74$\pm$ 2.74& 0.25$\pm$ 0.54& 3.10$\pm$ 1.78& 5.96$\pm$ 2.34& 2.85$\pm$ 1.61& 5.71$\pm$ 2.23& 2.86$\pm$ 1.19\\
PoC& 0.86$\pm$ 0.65& 1.23$\pm$ 1.03& 4.35$\pm$ 2.44& 6.81$\pm$ 2.95& 0.37$\pm$ 0.47& 3.49$\pm$ 1.88& 5.95$\pm$ 2.48& 3.12$\pm$ 1.73& 5.58$\pm$ 2.37& 2.46$\pm$ 0.94\\
PN?& 1.12$\pm$ 0.76& 1.75$\pm$ 1.19& 5.18$\pm$ 1.98& 8.15$\pm$ 2.59& 0.63$\pm$ 0.57& 4.06$\pm$ 1.70& 7.03$\pm$ 2.45& 3.42$\pm$ 1.57& 6.40$\pm$ 2.43& 2.98$\pm$ 1.16\\
bub& 0.99$\pm$ 0.68& 1.37$\pm$ 0.95& 4.66$\pm$ 2.12& 8.31$\pm$ 2.57& 0.38$\pm$ 0.38& 3.67$\pm$ 1.75& 7.32$\pm$ 2.32& 3.29$\pm$ 1.55& 6.94$\pm$ 2.15& 3.64$\pm$ 1.22\\
EmO& 0.99$\pm$ 0.70& 1.44$\pm$ 0.95& 4.41$\pm$ 2.15& 6.92$\pm$ 2.77& 0.45$\pm$ 0.39& 3.42$\pm$ 1.66& 5.93$\pm$ 2.37& 2.97$\pm$ 1.56& 5.47$\pm$ 2.31& 2.51$\pm$ 1.19\\
\end{tabular}
\caption{Average colours and magnitudes of SIMBAD object types} 
\label{simbadcolours5}

\end{table*}

\begin{table*}
\footnotesize\addtolength{\tabcolsep}{-4pt}
\begin{tabular}{cllllllllllllllllllllllllllll}

SIMBAD&K$_s$-W1&K$_s$-W2&K$_s$-W3&K$_s$-W4&W1-W2&W1-W3&W1-W4&W2-W3&W2-W4&W3-W4\\
\hline

RC*& 1.45$\pm$ 0.87& 2.48$\pm$ 1.18& 4.11$\pm$ 1.45& 4.90$\pm$ 1.61& 1.03$\pm$ 0.47& 2.66$\pm$ 0.95& 3.46$\pm$ 1.23& 1.63$\pm$ 0.67& 2.42$\pm$ 1.02& 0.79$\pm$ 0.46\\
Ro*& 0.11$\pm$ 0.10& 0.09$\pm$ 0.14& 0.22$\pm$ 0.34& 0.48$\pm$ 0.75&-0.02$\pm$ 0.09& 0.11$\pm$ 0.30& 0.38$\pm$ 0.73& 0.13$\pm$ 0.29& 0.39$\pm$ 0.72& 0.27$\pm$ 0.50\\
a2*& 0.02$\pm$ 0.09& 0.03$\pm$ 0.10&-0.04$\pm$ 0.14& 0.16$\pm$ 0.46& 0.00$\pm$ 0.10&-0.06$\pm$ 0.14& 0.14$\pm$ 0.46&-0.06$\pm$ 0.15& 0.13$\pm$ 0.47& 0.20$\pm$ 0.37\\
El*& 0.05$\pm$ 0.14& 0.08$\pm$ 0.14& 0.12$\pm$ 0.42& 0.62$\pm$ 1.27& 0.03$\pm$ 0.15& 0.07$\pm$ 0.40& 0.57$\pm$ 1.25& 0.04$\pm$ 0.39& 0.54$\pm$ 1.26& 0.50$\pm$ 0.97\\
Psr& 0.51$\pm$ 0.60& 0.53$\pm$ 0.88& 1.68$\pm$ 1.32& 3.25$\pm$ 2.02& 0.02$\pm$ 0.33& 1.17$\pm$ 1.30& 2.74$\pm$ 2.16& 1.15$\pm$ 1.38& 2.72$\pm$ 2.31& 1.57$\pm$ 1.12\\
BY*& 0.09$\pm$ 0.13& 0.12$\pm$ 0.15& 0.18$\pm$ 0.46& 0.40$\pm$ 0.93& 0.04$\pm$ 0.14& 0.09$\pm$ 0.42& 0.32$\pm$ 0.89& 0.05$\pm$ 0.37& 0.28$\pm$ 0.85& 0.23$\pm$ 0.52\\
RS*& 0.10$\pm$ 0.18& 0.15$\pm$ 0.18& 0.17$\pm$ 0.35& 0.32$\pm$ 0.53& 0.05$\pm$ 0.15& 0.08$\pm$ 0.29& 0.22$\pm$ 0.46& 0.03$\pm$ 0.23& 0.17$\pm$ 0.42& 0.14$\pm$ 0.25\\
\end{tabular}

\caption{Average colours and magnitudes of SIMBAD object types} 
\label{simbadcolours6}

\end{table*}

\begin{table*}
\footnotesize\addtolength{\tabcolsep}{-4pt}
\begin{tabular}{cllllllllllllllllllllllllllll}

SIMBAD&J&H&K$_s$&W1&W2&W3&W4&\textit{ext}\\
\hline
Rad&15.56$\pm$ 1.29&14.74$\pm$ 1.36&14.11$\pm$ 1.34&13.24$\pm$ 1.40&12.68$\pm$ 1.38& 9.31$\pm$ 1.38& 6.89$\pm$ 1.54&0.94$\pm$1.70\\
 cm&15.81$\pm$ 1.02&15.06$\pm$ 0.97&14.39$\pm$ 0.94&13.31$\pm$ 1.06&12.48$\pm$ 1.04& 9.84$\pm$ 1.28& 7.51$\pm$ 1.38&0.59$\pm$1.35\\
 mm&13.67$\pm$ 1.83&11.84$\pm$ 1.70&10.96$\pm$ 1.75& 9.84$\pm$ 1.57& 9.36$\pm$ 1.62& 6.57$\pm$ 1.94& 3.47$\pm$ 2.26&1.32$\pm$1.09\\
smm&14.65$\pm$ 1.72&12.89$\pm$ 1.84&11.89$\pm$ 2.08&10.68$\pm$ 2.15&10.02$\pm$ 2.33& 7.09$\pm$ 2.20& 4.07$\pm$ 2.47&1.65$\pm$1.55\\
 HI&15.21$\pm$ 1.21&14.57$\pm$ 1.21&14.18$\pm$ 1.17&13.36$\pm$ 1.05&13.21$\pm$ 1.00& 9.60$\pm$ 0.53& 7.38$\pm$ 0.60&1.73$\pm$1.99\\
Mas&12.68$\pm$ 2.01&10.43$\pm$ 1.97& 9.02$\pm$ 2.33& 7.38$\pm$ 2.44& 6.26$\pm$ 2.82& 3.42$\pm$ 2.49& 0.96$\pm$ 2.14&1.32$\pm$1.52\\
 IR&12.13$\pm$ 3.15&10.82$\pm$ 3.29&10.09$\pm$ 3.38& 9.29$\pm$ 3.32& 8.93$\pm$ 3.43& 6.81$\pm$ 2.65& 5.02$\pm$ 2.44&0.89$\pm$1.31\\
\end{tabular}

\caption{Average colours and magnitudes of SIMBAD object types} 
\label{simbadcolours7}

\end{table*}

\begin{table*}
\footnotesize\addtolength{\tabcolsep}{-4pt}
\begin{tabular}{cllllllllllllllllllllllllllll}

SIMBAD&J&H&K$_s$&W1&W2&W3&W4&\textit{ext}\\
\hline
LXB&13.39$\pm$ 3.47&12.57$\pm$ 3.61&12.27$\pm$ 3.63&10.57$\pm$ 2.72&10.65$\pm$ 3.04& 8.48$\pm$ 2.14& 6.72$\pm$ 1.32&1.25$\pm$1.26\\
HXB&10.96$\pm$ 2.78&10.36$\pm$ 2.62& 9.94$\pm$ 2.59& 9.28$\pm$ 2.38& 8.99$\pm$ 2.41& 7.65$\pm$ 2.14& 6.07$\pm$ 2.29&0.88$\pm$1.29\\
ISM&13.20$\pm$ 2.00&12.44$\pm$ 1.87&12.03$\pm$ 1.91&11.26$\pm$ 1.88&11.01$\pm$ 2.14& 8.15$\pm$ 2.11& 5.30$\pm$ 2.42&1.27$\pm$1.18\\
PoC&12.93$\pm$ 2.85&11.67$\pm$ 2.49&10.99$\pm$ 2.31&10.13$\pm$ 1.98& 9.76$\pm$ 1.82& 6.64$\pm$ 2.01& 4.18$\pm$ 2.26&1.20$\pm$1.03\\
PN?&13.58$\pm$ 1.83&12.62$\pm$ 1.90&11.84$\pm$ 1.91&10.72$\pm$ 2.00&10.09$\pm$ 2.19& 6.66$\pm$ 2.19& 3.69$\pm$ 2.58&0.80$\pm$0.77\\
bub&13.40$\pm$ 1.75&11.79$\pm$ 1.67&11.01$\pm$ 1.75&10.02$\pm$ 1.60& 9.64$\pm$ 1.63& 6.34$\pm$ 1.70& 2.70$\pm$ 2.11&0.94$\pm$0.24\\
EmO&13.83$\pm$ 2.17&13.09$\pm$ 2.21&12.54$\pm$ 2.28&11.55$\pm$ 2.19&11.10$\pm$ 2.30& 8.13$\pm$ 1.97& 5.62$\pm$ 2.29&1.18$\pm$1.44\\
\end{tabular}

\caption{Average colours and magnitudes of SIMBAD object types} 
\label{simbadcolours8}

\end{table*}

\begin{table*}
\footnotesize\addtolength{\tabcolsep}{-4pt}
\begin{tabular}{cllllllllllllllllllllllllllll}

SIMBAD&J&H&K$_s$&W1&W2&W3&W4&\textit{ext}\\
\hline
RC*&10.77$\pm$ 2.41& 9.81$\pm$ 2.17& 8.92$\pm$ 2.00& 7.47$\pm$ 1.89& 6.44$\pm$ 1.95& 4.81$\pm$ 1.80& 4.02$\pm$ 1.85&0.48$\pm$0.52\\
Ro*& 8.82$\pm$ 1.19& 8.29$\pm$ 1.17& 8.13$\pm$ 1.18& 8.02$\pm$ 1.16& 8.03$\pm$ 1.17& 7.91$\pm$ 1.05& 7.64$\pm$ 0.93&0.10$\pm$0.31\\
a2*& 6.96$\pm$ 1.06& 6.98$\pm$ 1.04& 6.96$\pm$ 1.04& 6.94$\pm$ 1.03& 6.94$\pm$ 1.09& 7.00$\pm$ 1.00& 6.80$\pm$ 1.00&0.12$\pm$0.33\\
El*& 7.23$\pm$ 1.63& 7.13$\pm$ 1.55& 7.08$\pm$ 1.54& 7.03$\pm$ 1.53& 7.00$\pm$ 1.59& 6.96$\pm$ 1.42& 6.46$\pm$ 1.20&0.22$\pm$0.42\\
Psr&11.60$\pm$ 2.09&10.79$\pm$ 1.96&10.47$\pm$ 2.00& 9.96$\pm$ 2.20& 9.94$\pm$ 2.40& 8.79$\pm$ 1.77& 7.22$\pm$ 1.35&0.63$\pm$0.76\\
BY*& 7.83$\pm$ 1.46& 7.36$\pm$ 1.40& 7.22$\pm$ 1.38& 7.14$\pm$ 1.35& 7.10$\pm$ 1.38& 7.05$\pm$ 1.25& 6.82$\pm$ 1.19&0.15$\pm$0.38\\
RS*& 7.67$\pm$ 1.40& 7.17$\pm$ 1.38& 7.03$\pm$ 1.39& 6.93$\pm$ 1.37& 6.88$\pm$ 1.43& 6.85$\pm$ 1.36& 6.71$\pm$ 1.33&0.14$\pm$0.34\\
\end{tabular}

\caption{Average colours and magnitudes of SIMBAD object types} 
\label{simbadcolours9}

\end{table*}

\newpage
\section{Real and spurious source separation}
\begin{table*}
\begin{tabular}{cc}
 \includegraphics[width=0.35\textwidth,trim=0cm 0cm 0cm 4cm]{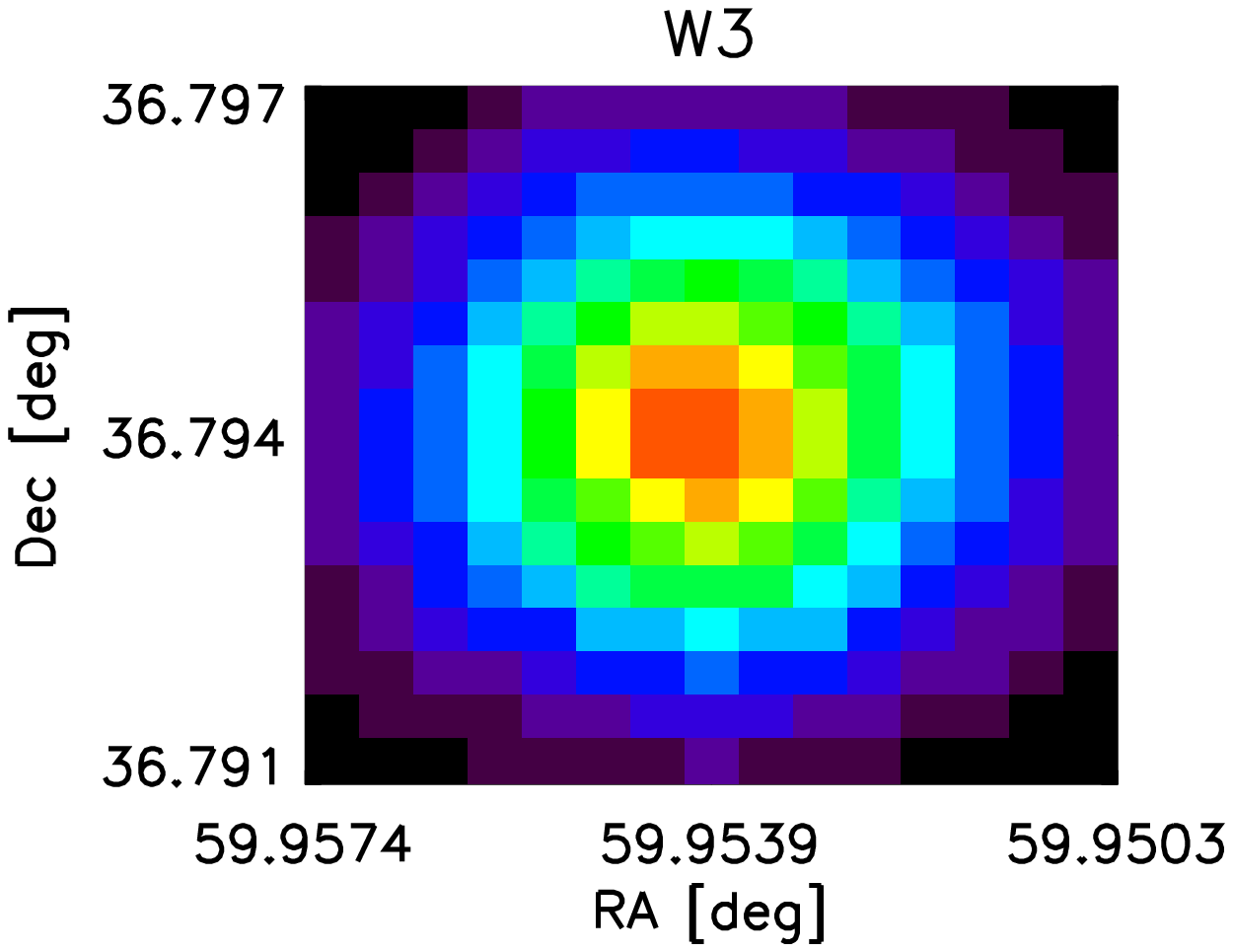} &  \includegraphics[width=0.35\textwidth,trim=0cm 0cm 0cm 4cm]{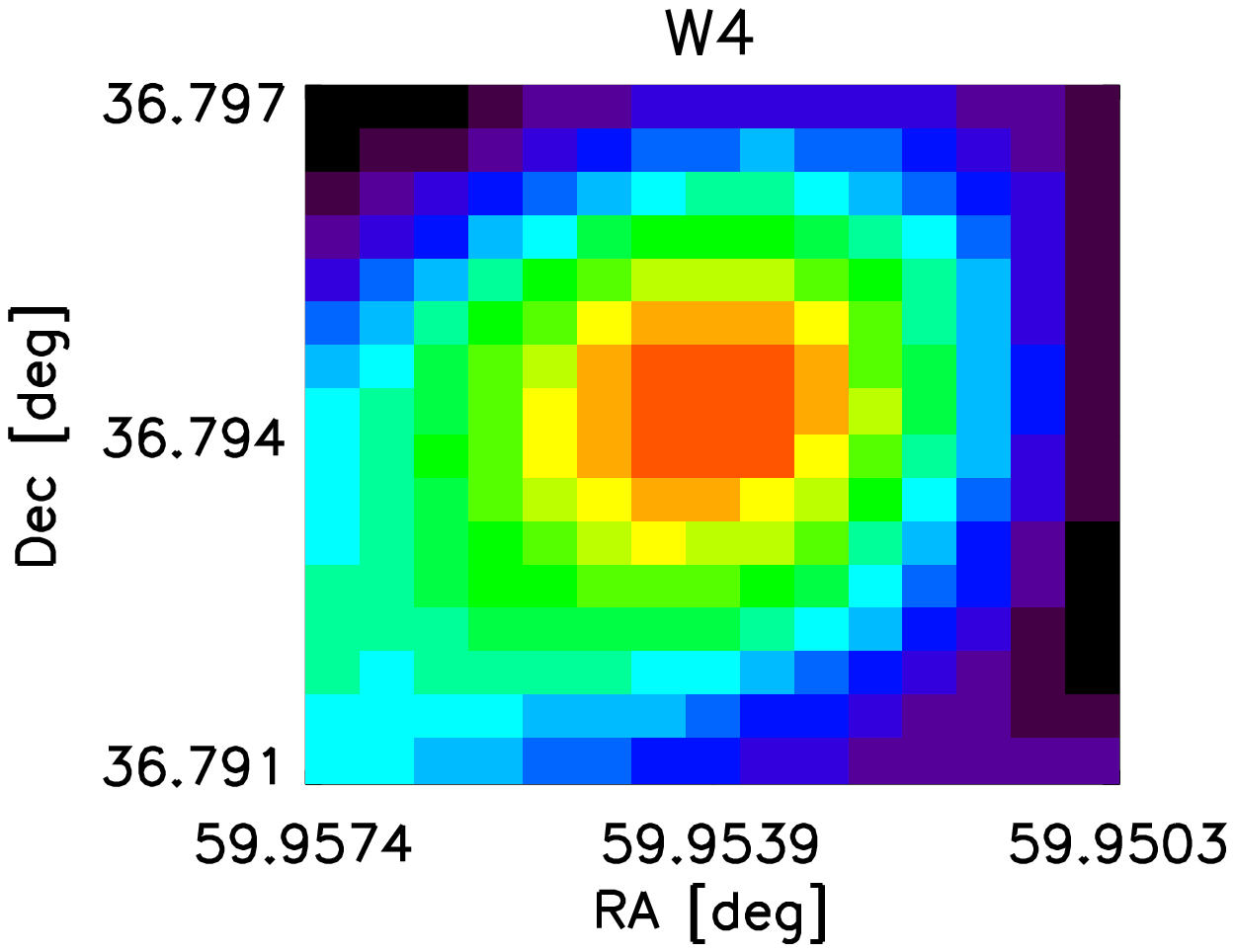}\\
 \\
 \includegraphics[width=0.35\textwidth,trim=0cm 0cm 0cm 4cm]{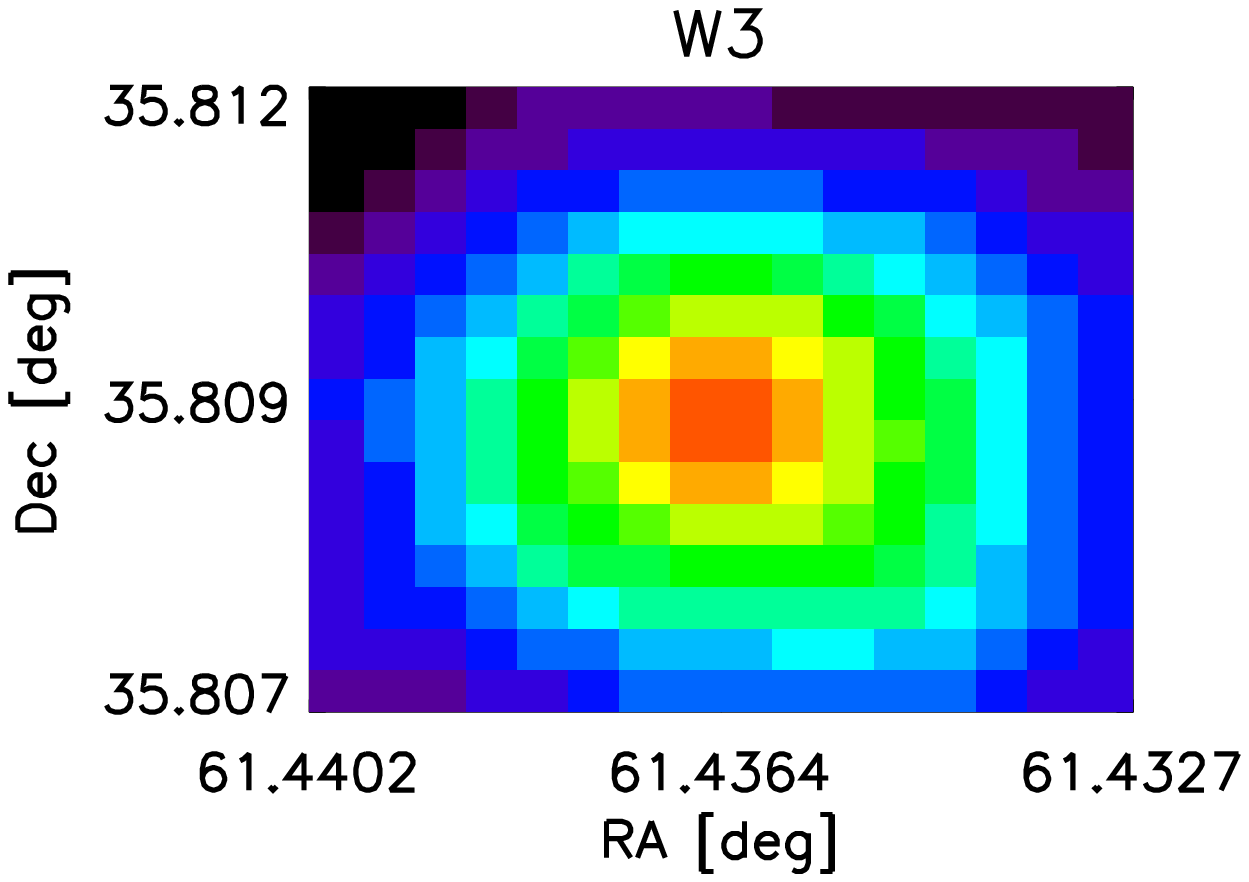} &  \includegraphics[width=0.35\textwidth,trim=0cm 0cm 0cm 4cm]{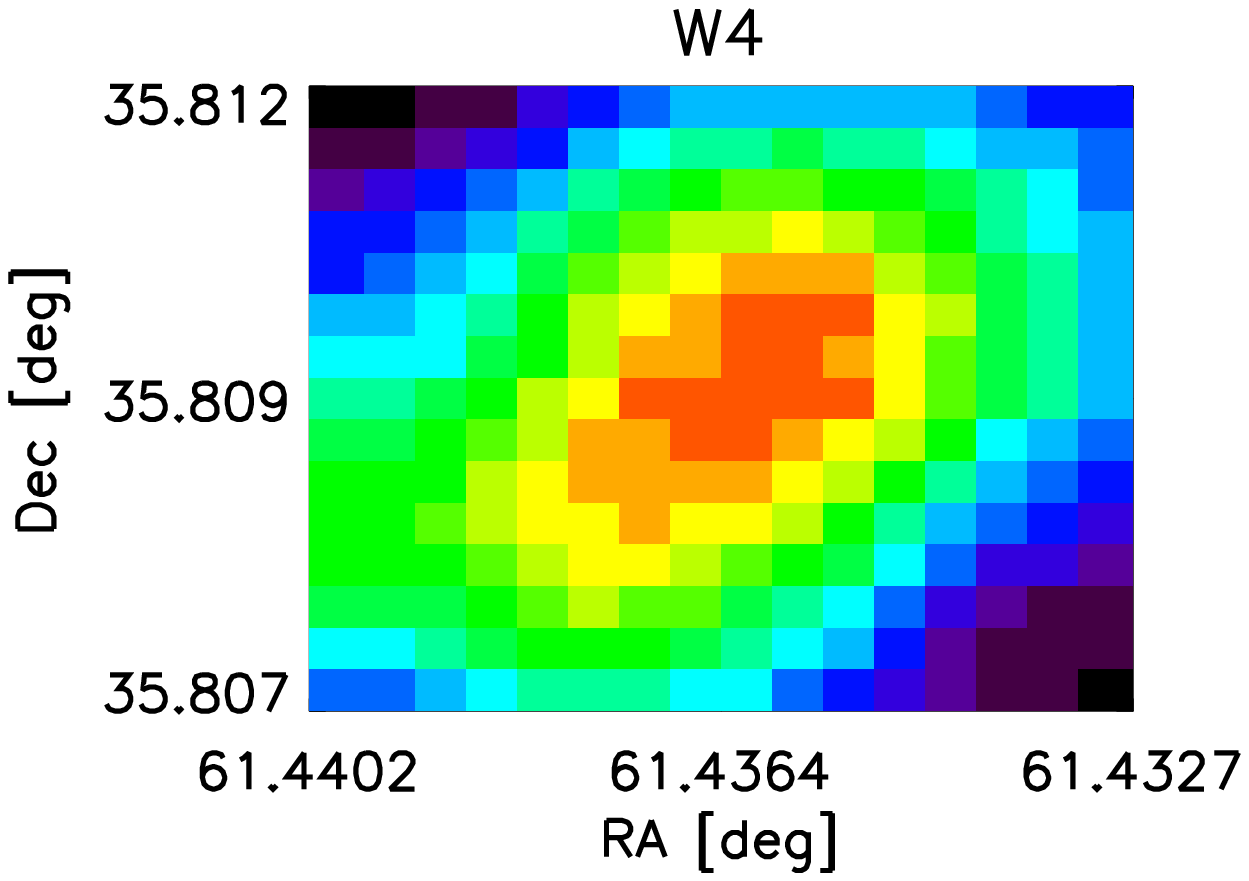}\\
 \\
 \includegraphics[width=0.35\textwidth,trim=0cm 0cm 0cm 4cm]{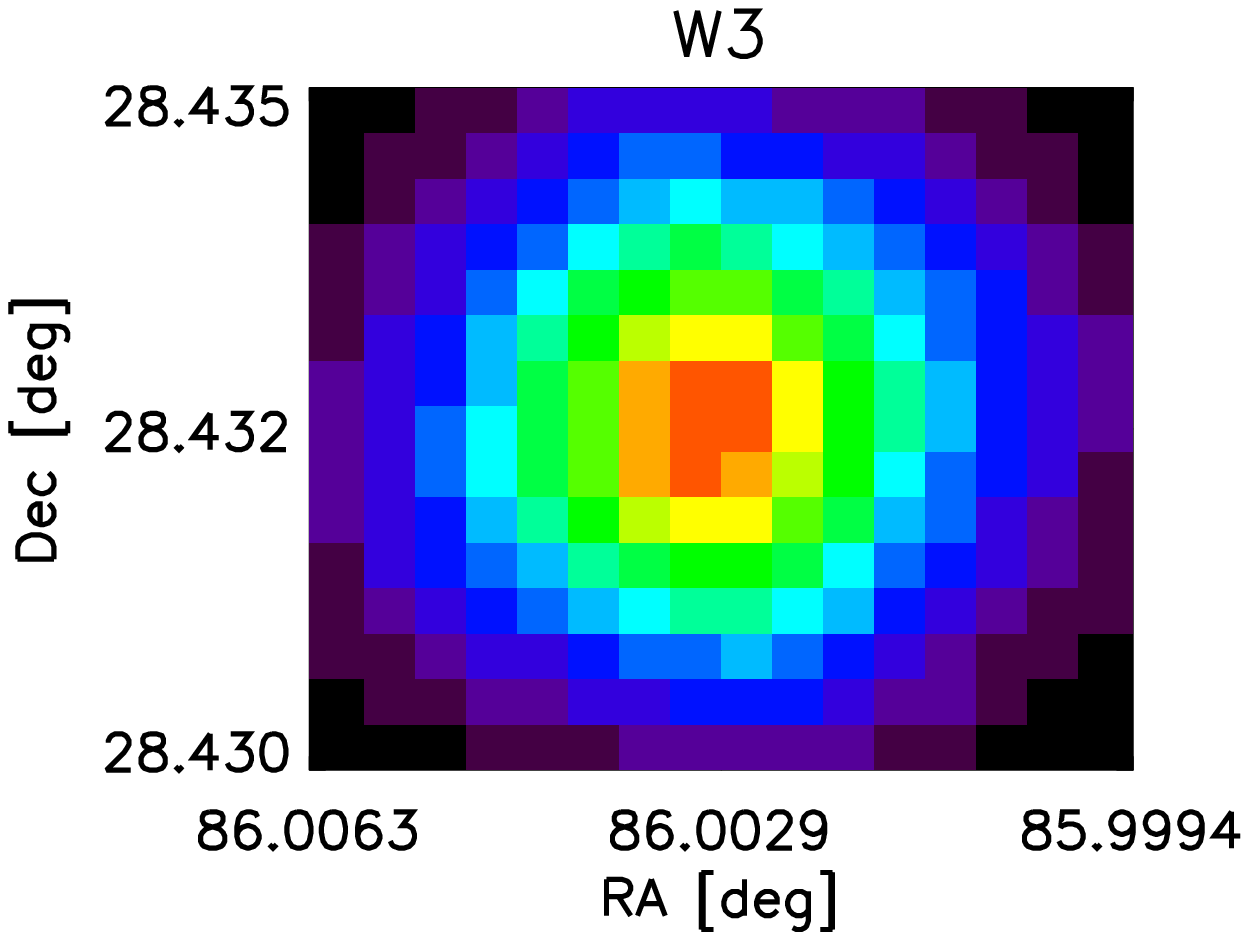} &  \includegraphics[width=0.35\textwidth,trim=0cm 0cm 0cm 4cm]{0857p287_ac51-w3-int-3_ra86.003192_dec28.432336_asec20.000.fits.ps}\\
 \\
 \includegraphics[width=0.35\textwidth,trim=0cm 0cm 0cm 4cm]{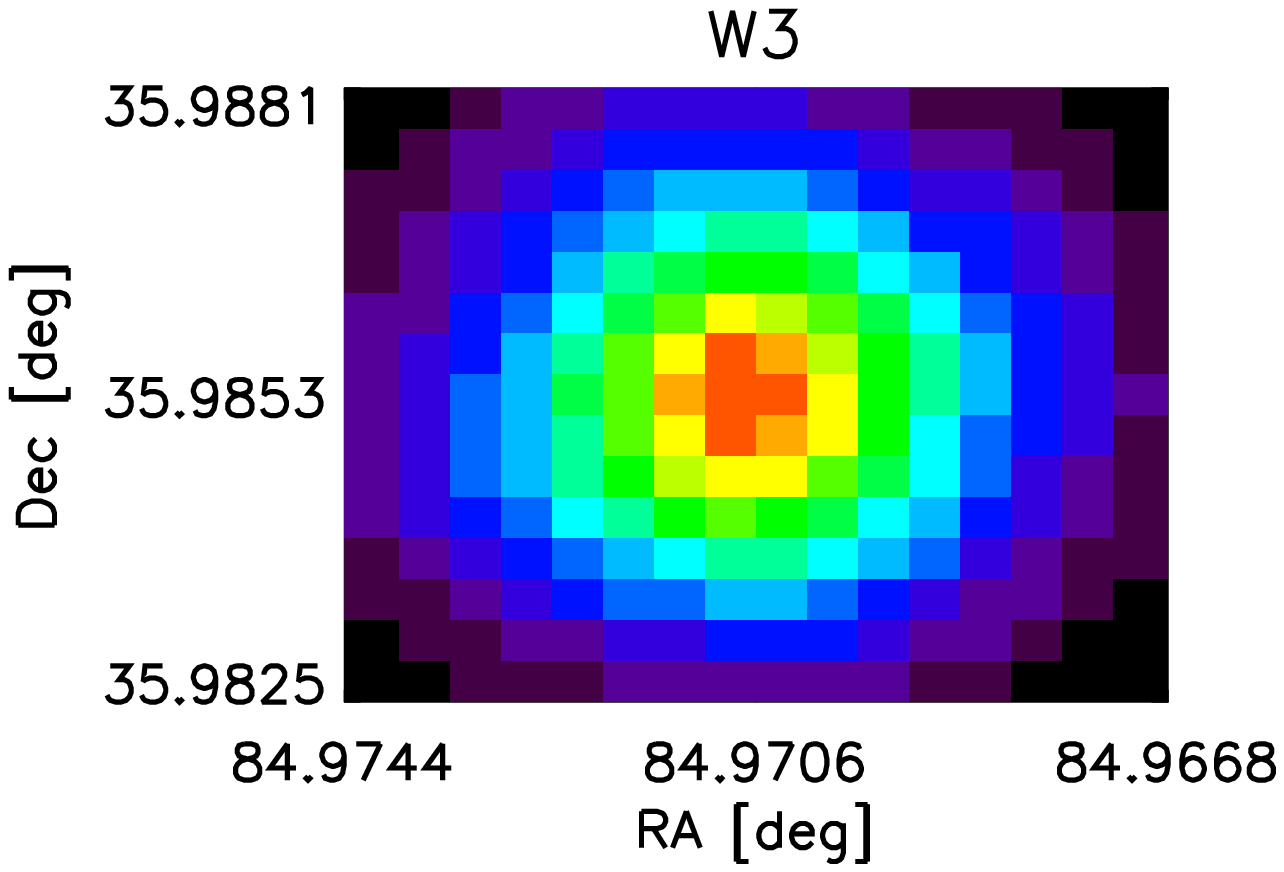} &  \includegraphics[width=0.35\textwidth,trim=0cm 0cm 0cm 4cm]{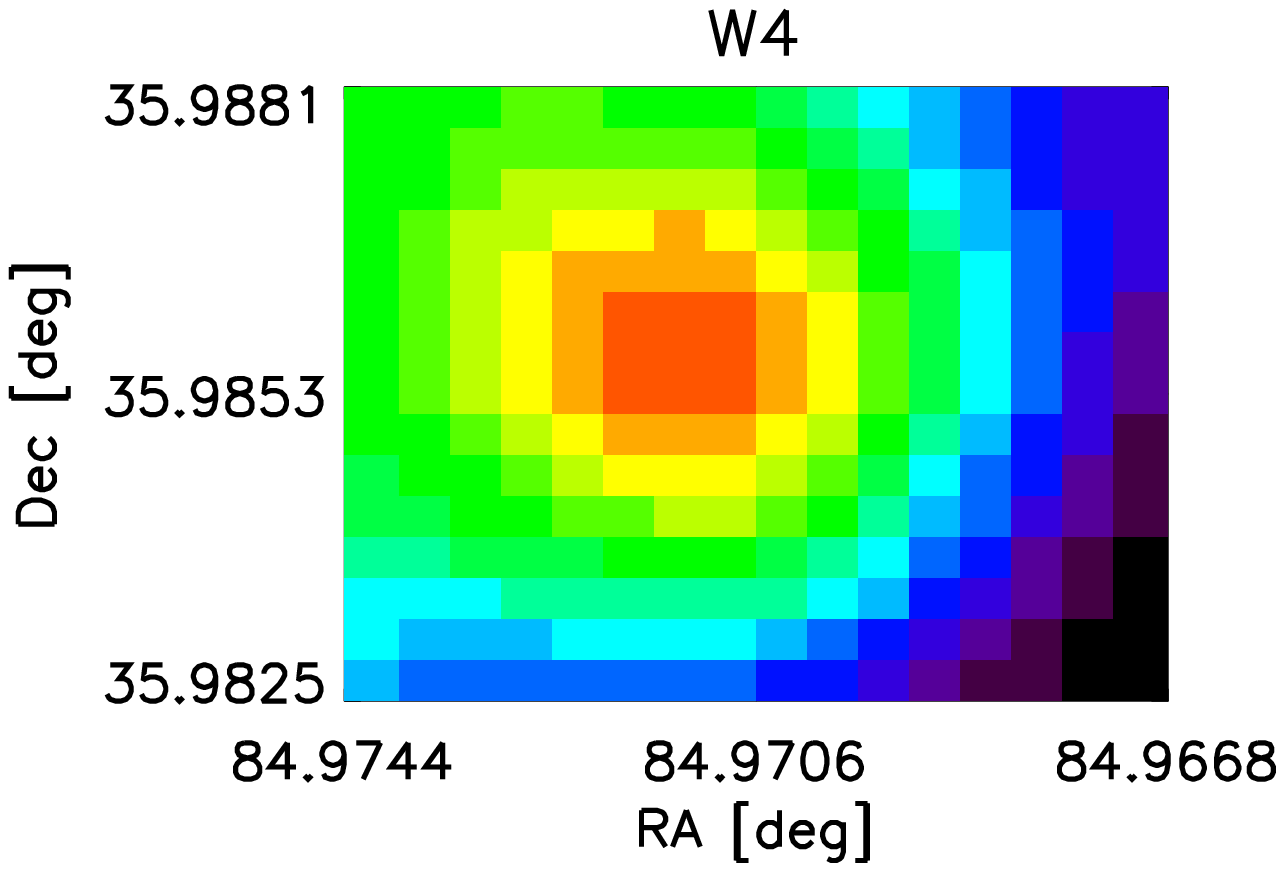}\\
 \\
 \includegraphics[width=0.35\textwidth,trim=0cm 0cm 0cm 3.5cm]{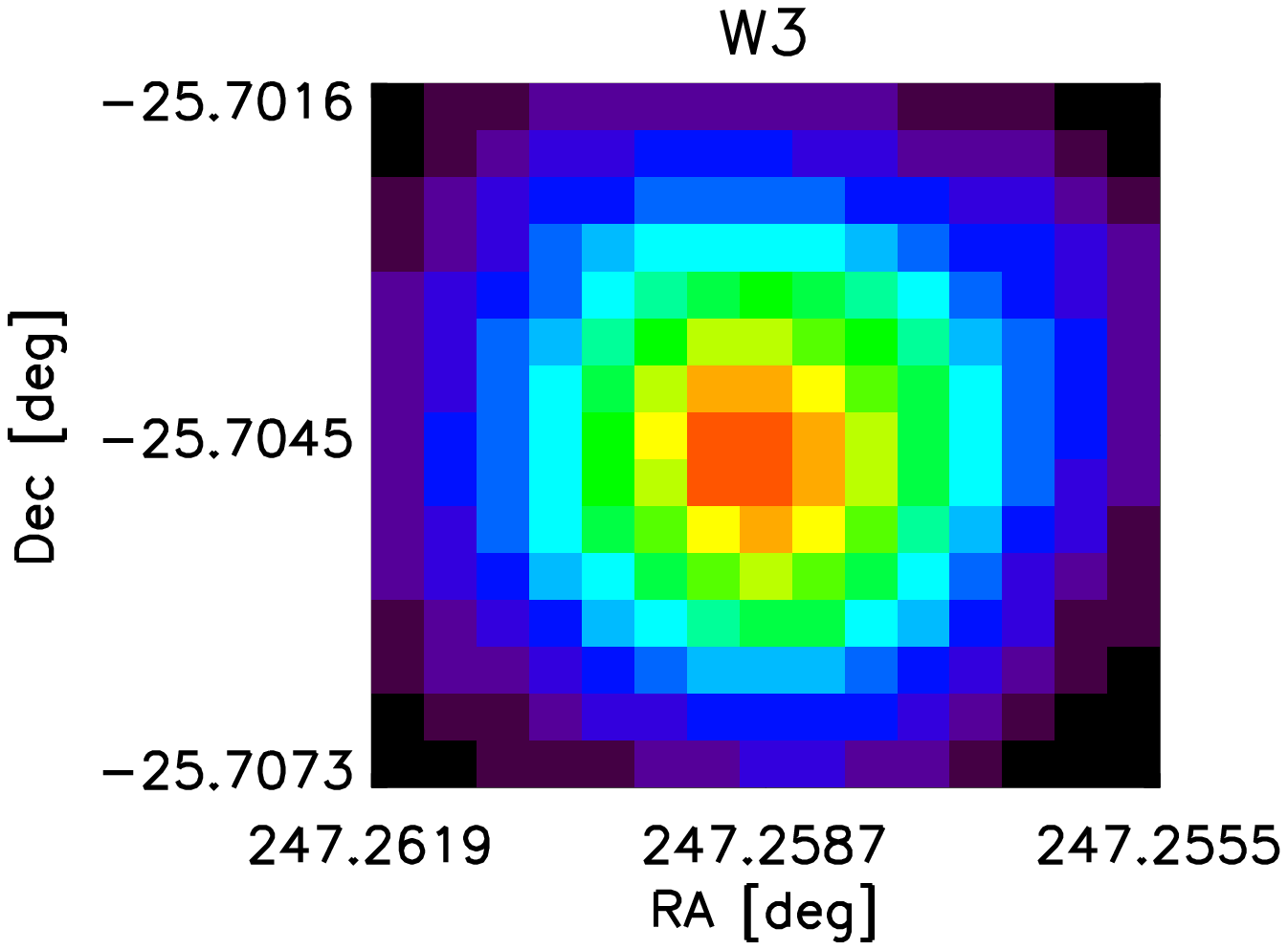} &  \includegraphics[width=0.35\textwidth,trim=0cm 0cm 0cm 3.5cm]{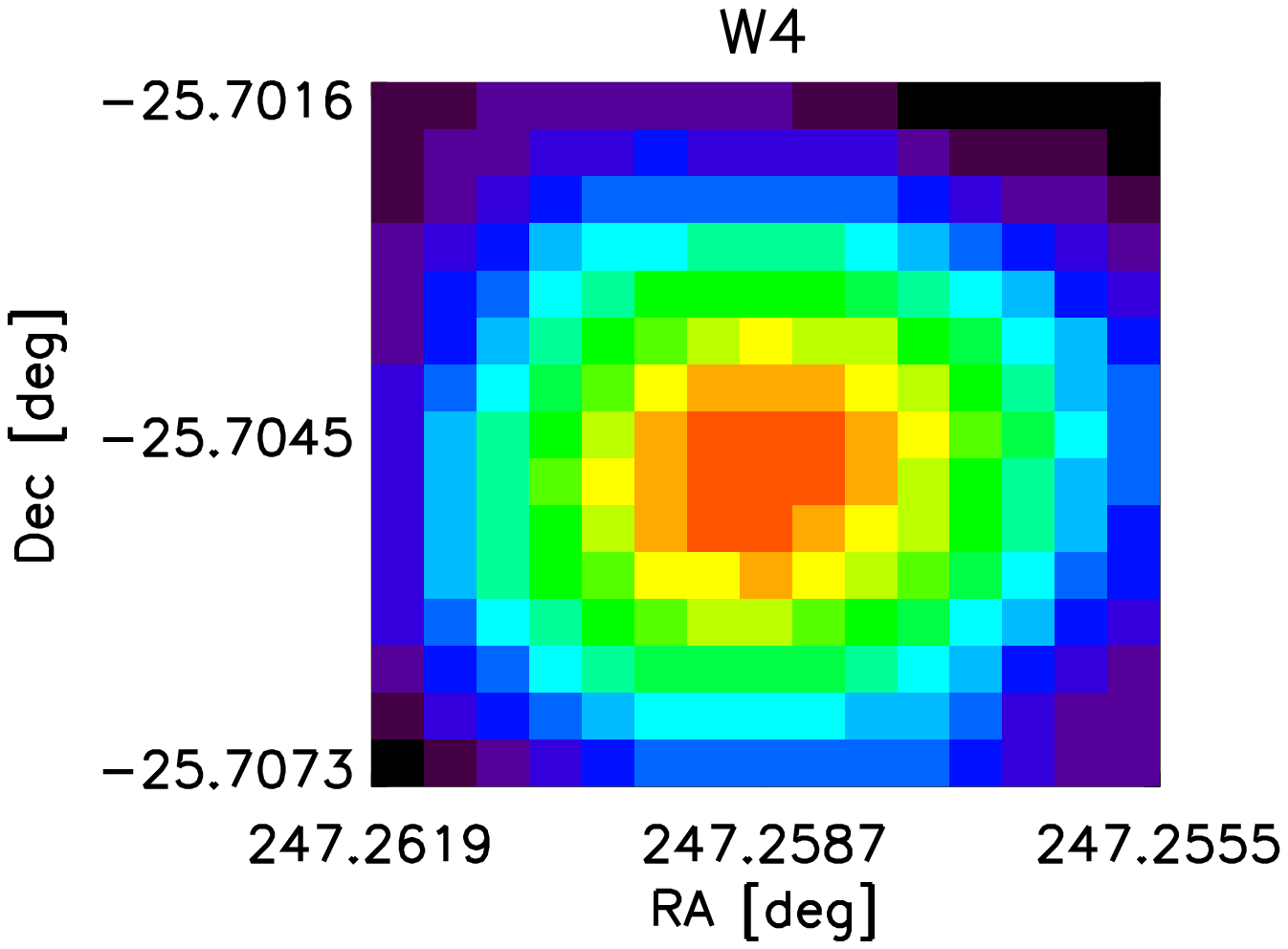}\\

\end{tabular}
\caption{Example of sources used as real for the spurious source identification training sample. Left column shows the W3 image, right column is for the W4 image. Stamps are centered on the catalogued AllWISE positions and cover 20$^{\prime\prime}\times$20$^{\prime\prime}$ size regions.}\label{realw3w4}
\end{table*}

\begin{table*}
\begin{tabular}{cc}
 \includegraphics[width=0.35\textwidth,trim=0cm 0cm 0cm 4cm]{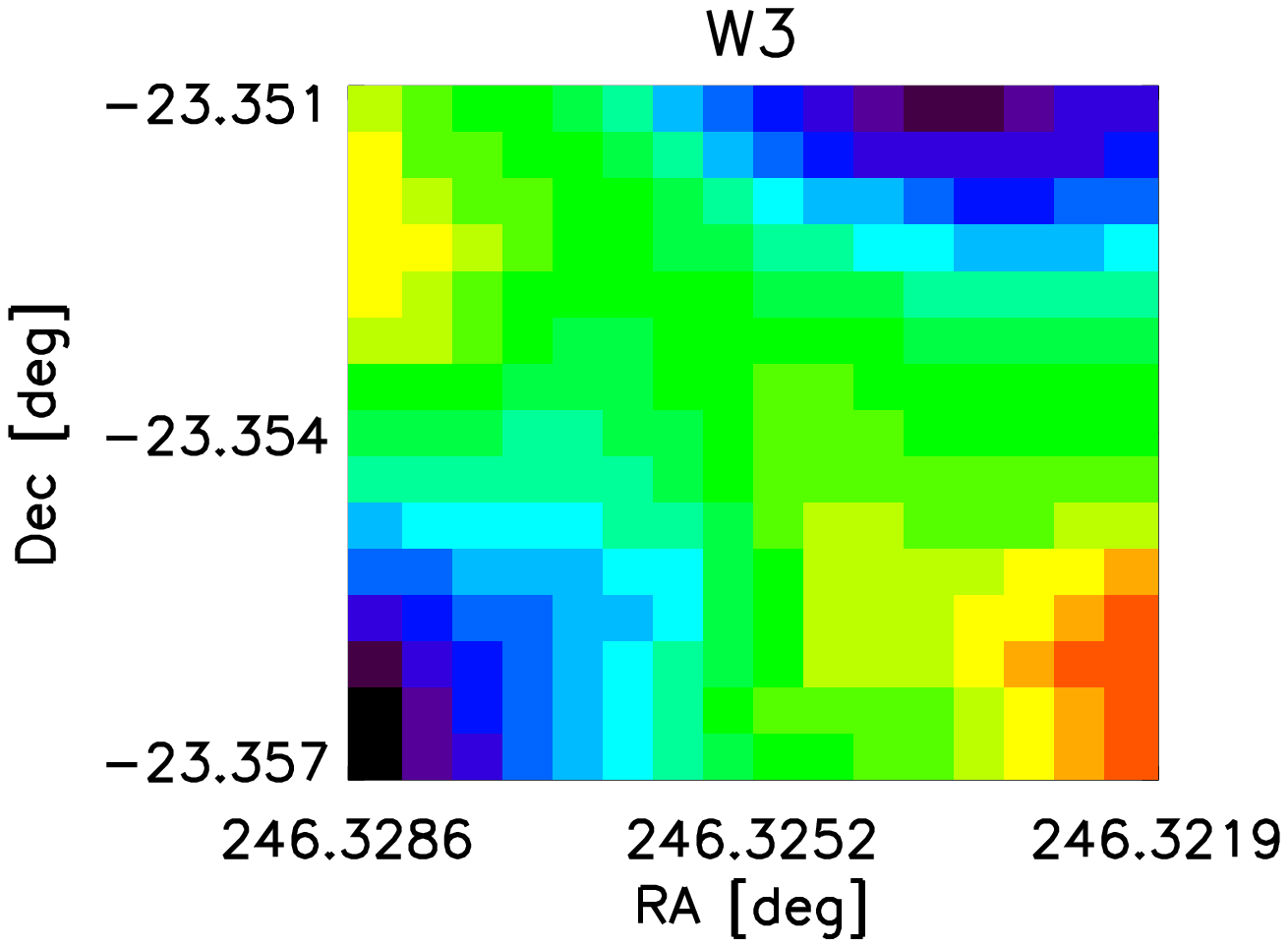} &  \includegraphics[width=0.35\textwidth,trim=0cm 0cm 0cm 4cm]{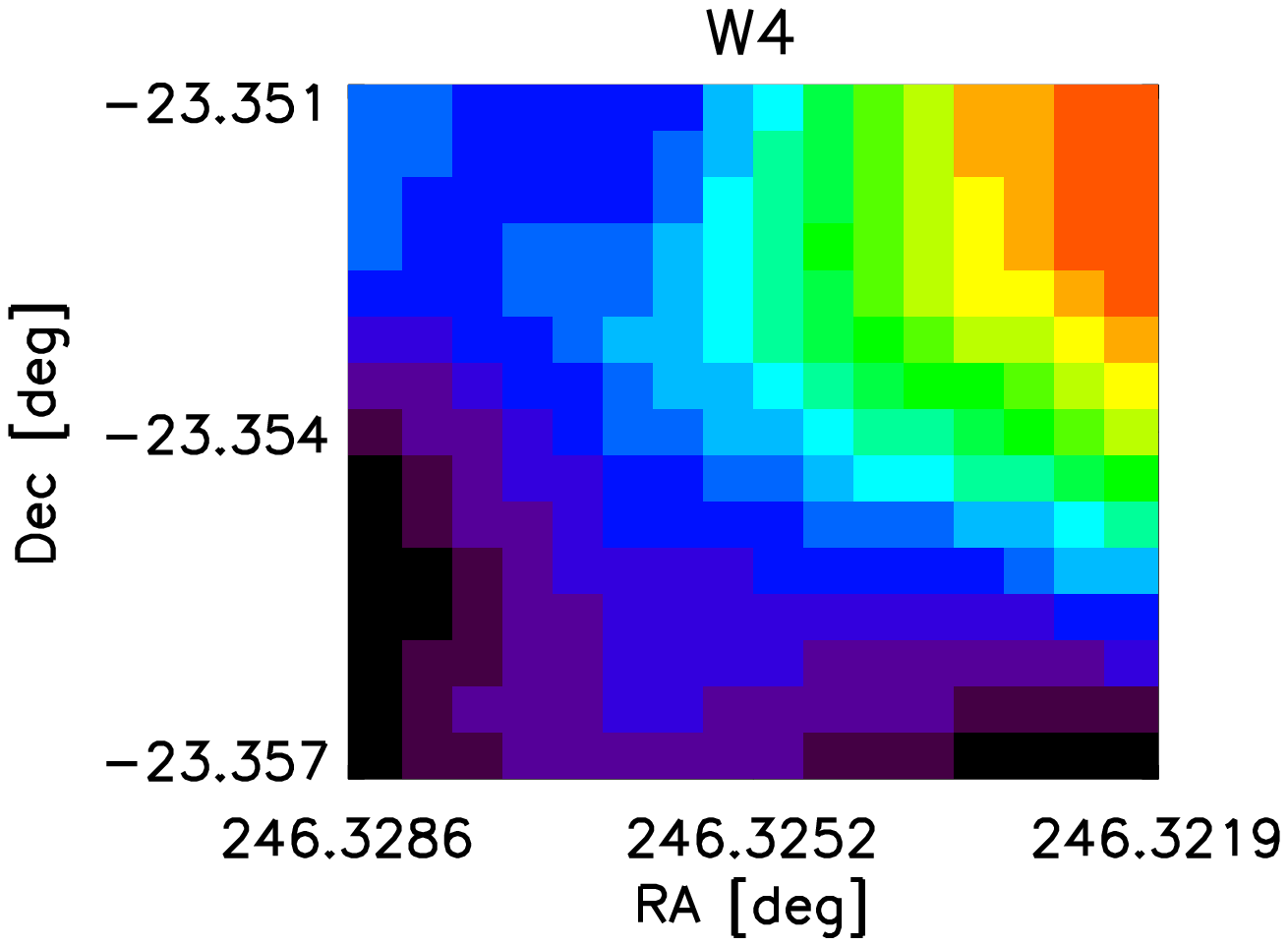}\\
 \\
 \includegraphics[width=0.35\textwidth,trim=0cm 0cm 0cm 3.5cm]{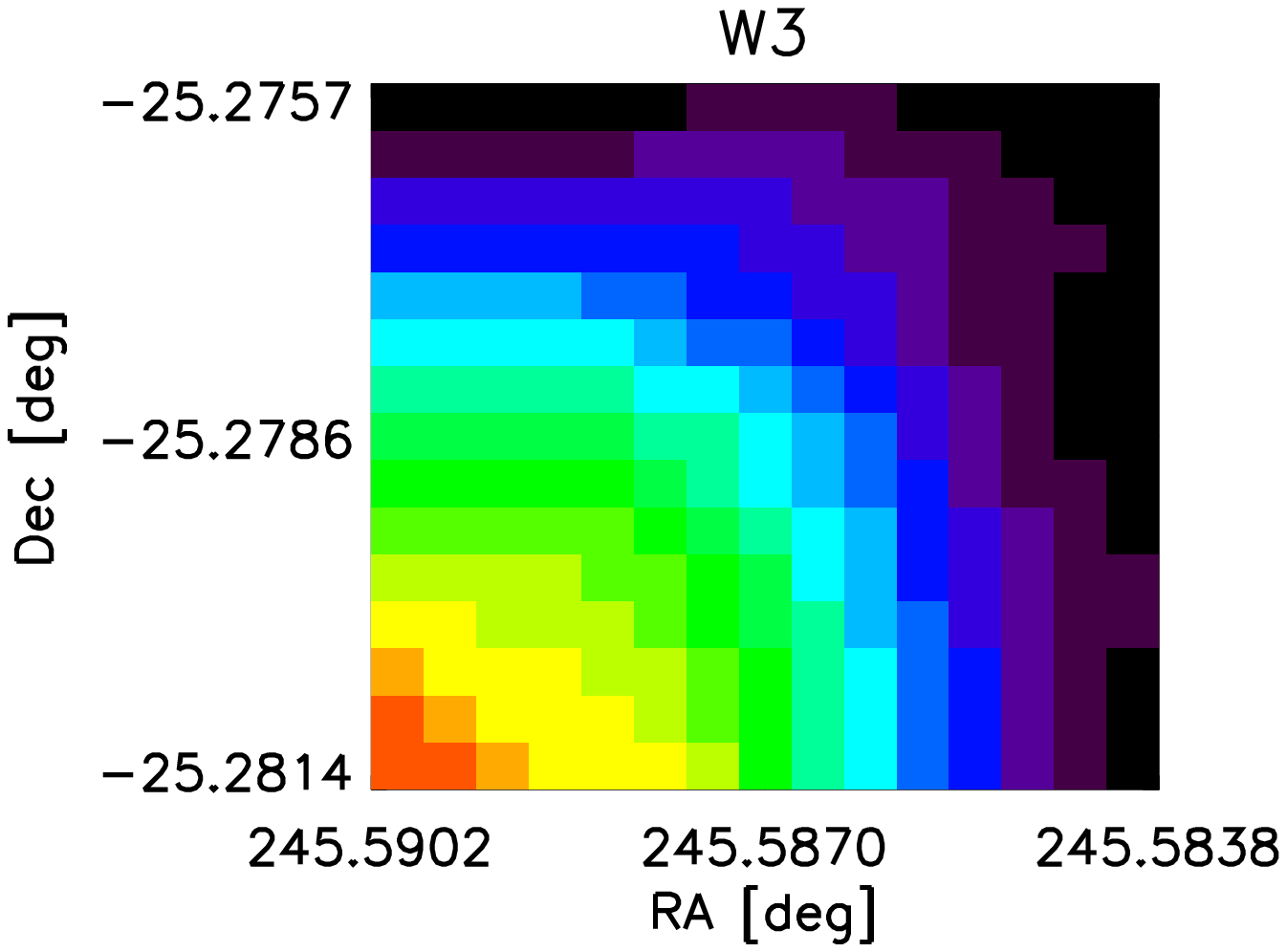} &  \includegraphics[width=0.35\textwidth,trim=0cm 0cm 0cm 3.5cm]{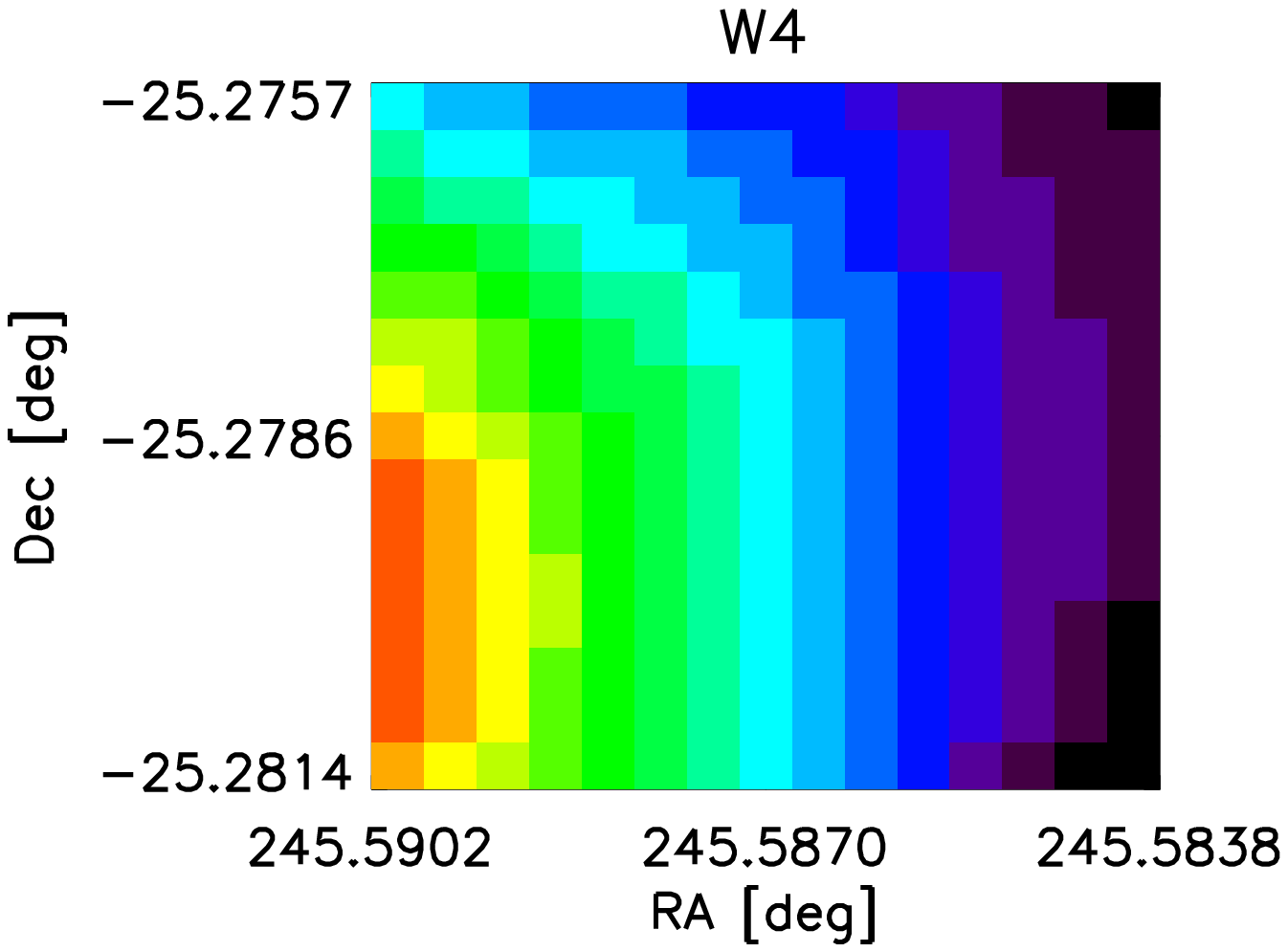}\\
 \\
 \includegraphics[width=0.35\textwidth,trim=0cm 0cm 0cm 4cm]{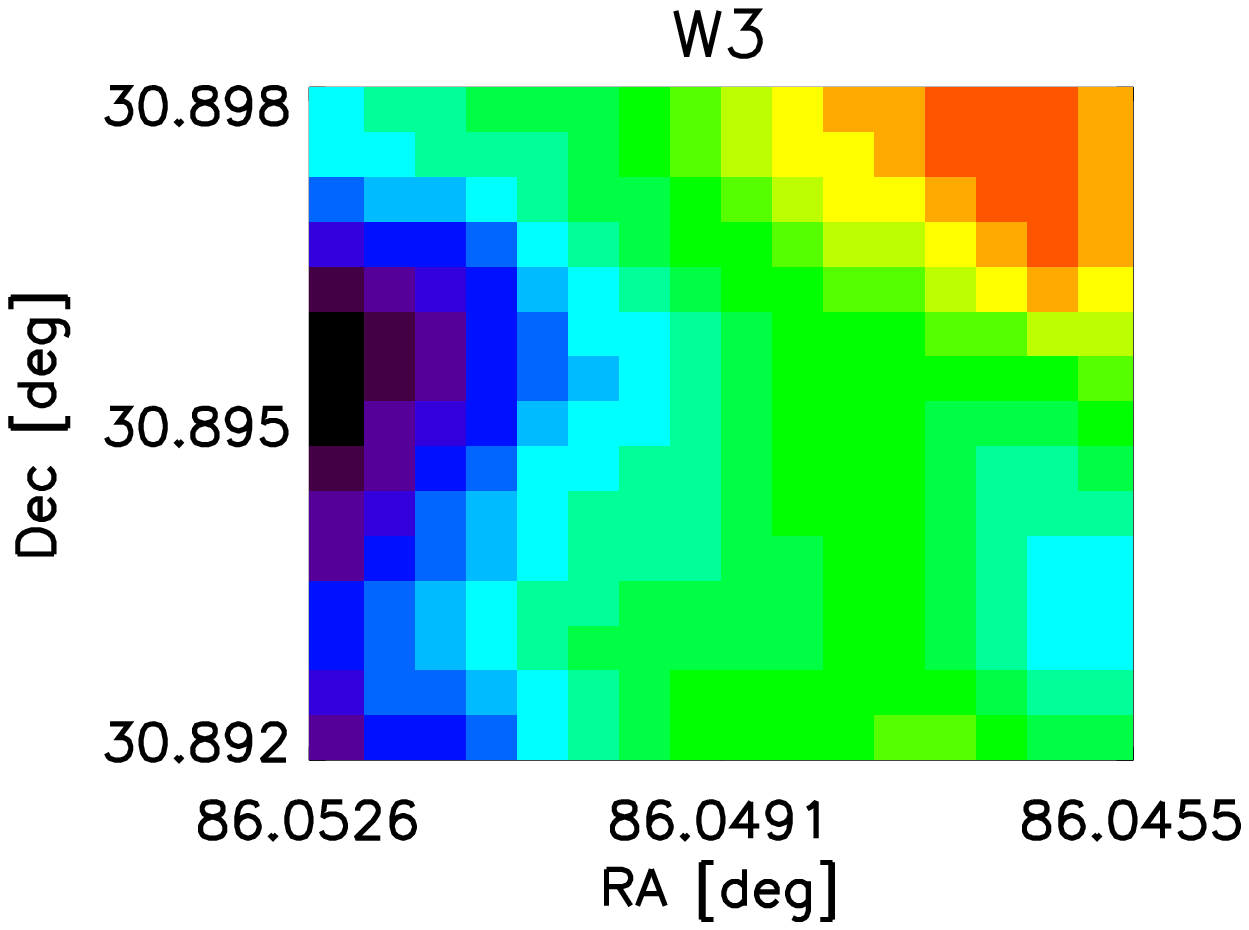} &  \includegraphics[width=0.35\textwidth,trim=0cm 0cm 0cm 4cm]{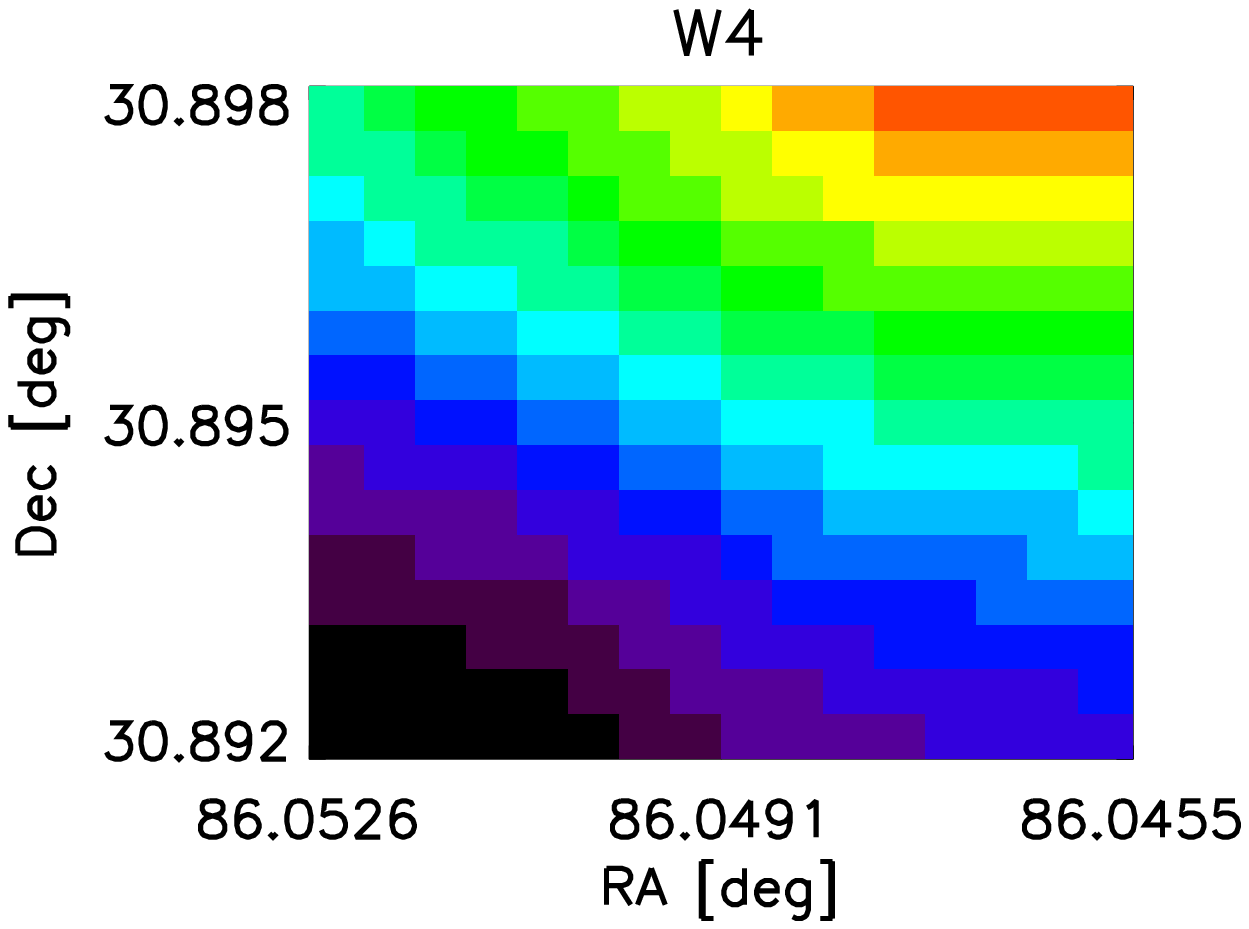}\\
 \\
 \includegraphics[width=0.35\textwidth,trim=0cm 0cm 0cm 4cm]{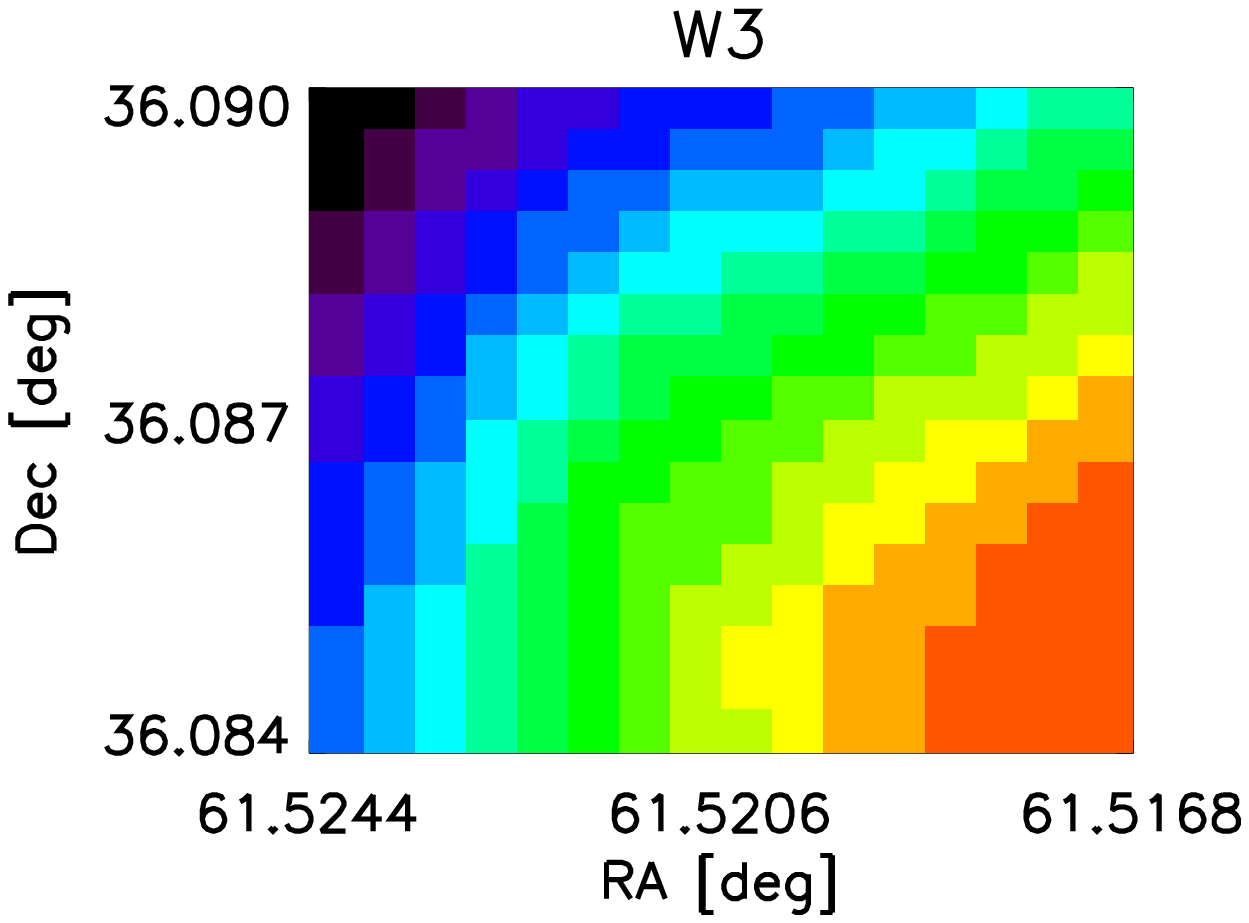} &  \includegraphics[width=0.35\textwidth,trim=0cm 0cm 0cm 4cm]{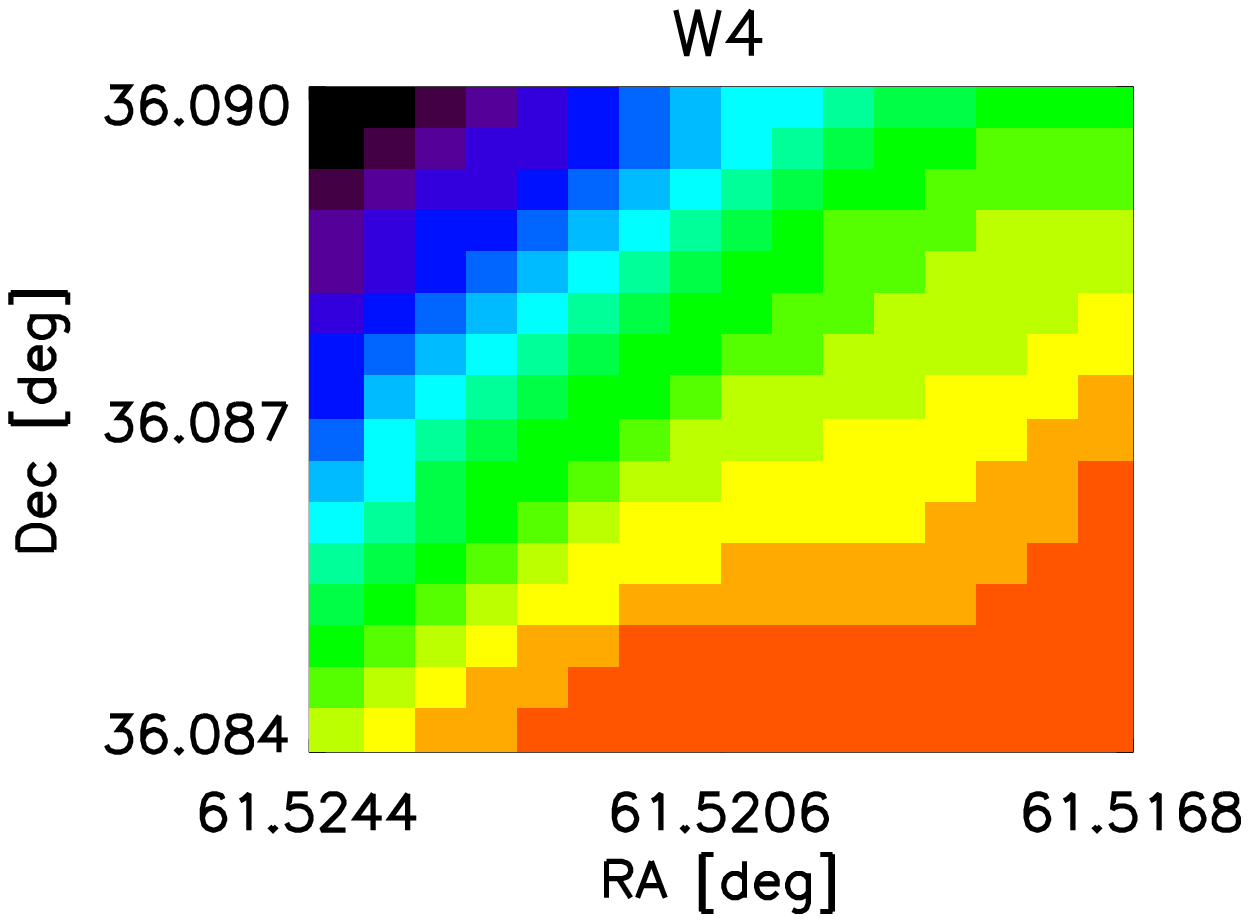}\\
 \\
 \includegraphics[width=0.35\textwidth,trim=0cm 0cm 0cm 3.5cm]{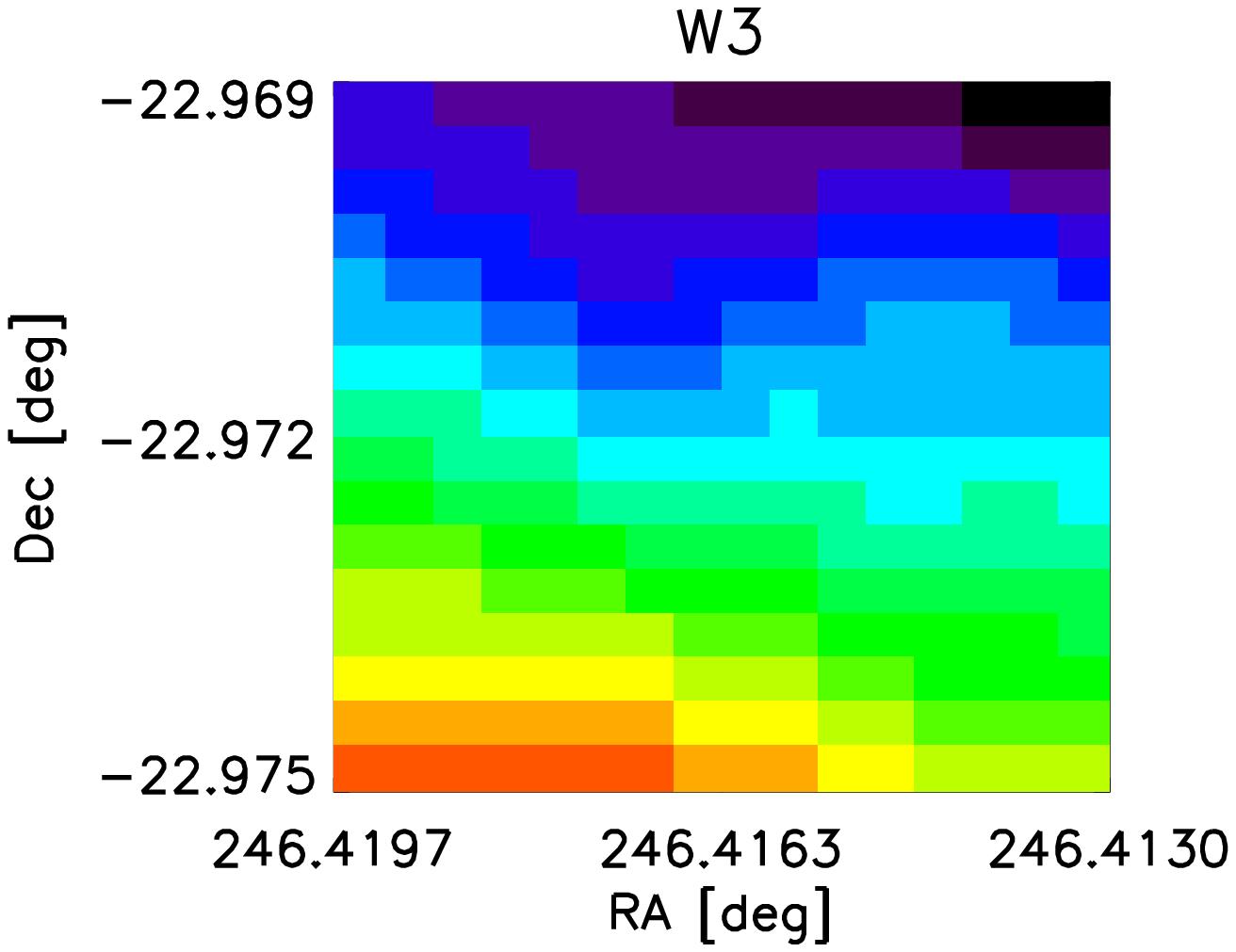} &  \includegraphics[width=0.35\textwidth,trim=0cm 0cm 0cm 3.5cm]{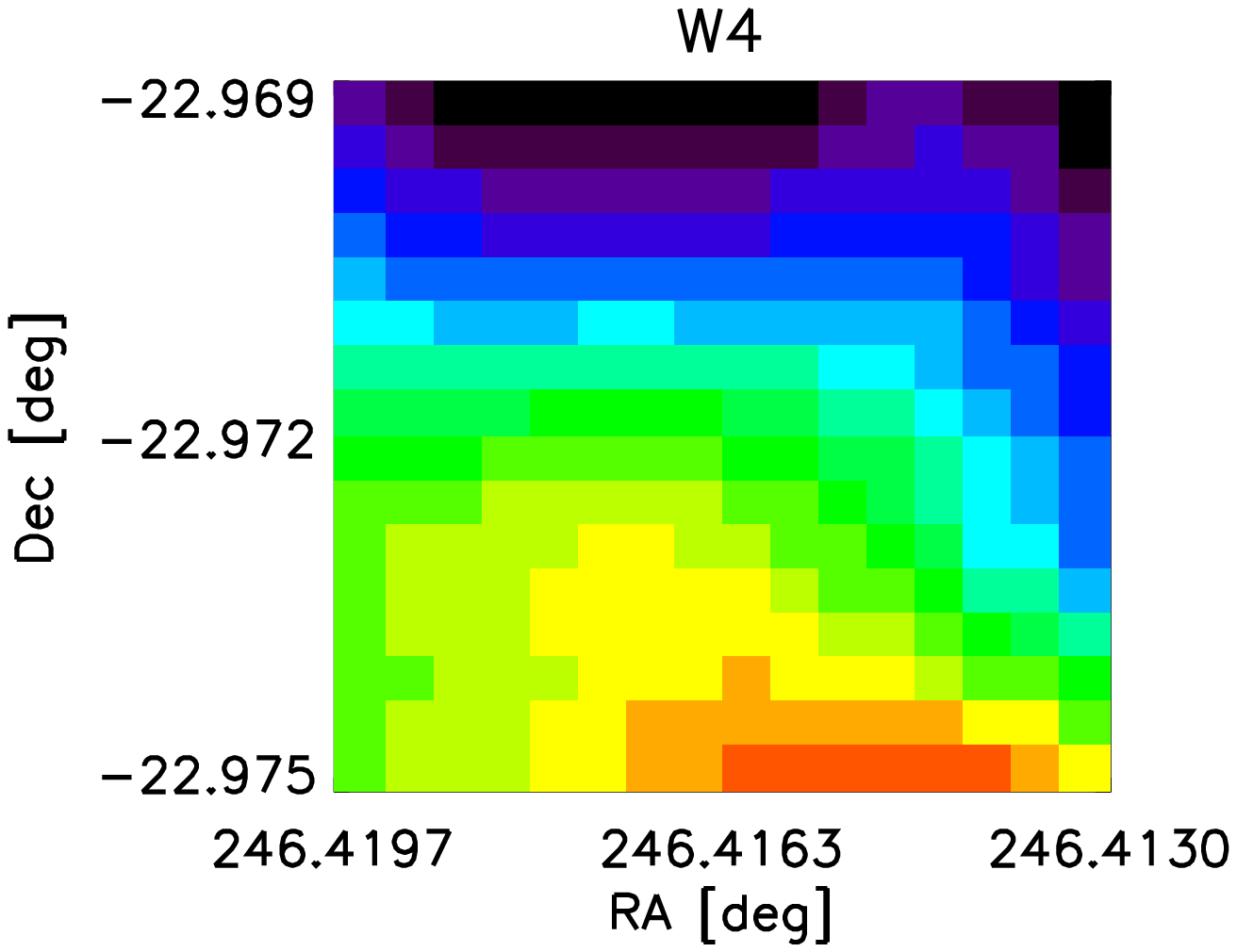}\\

\end{tabular}
\caption{Example of sources selected as spurious for the spurious source identification training sample. Left column shows the W3 image, right column is for the W4 image. Stamps are centered on the catalogued AllWISE positions and cover 20$^{\prime\prime}\times$20$^{\prime\prime}$ size regions.}\label{fakew3w4}
\end{table*}

\newpage
\section{Grouping of SIMBAD objects}

SVM is a powerful algorithm to separate a number of objects classes in the multi-dimensional data space. However, the SIMBAD database list 235 object types. To have statistically more robust training samples and to simplify the calculations, we created 11 sub-types from the most relevant object types. These types are listed in Section \ref{svm}. Here we present their colours and the \textit{ext} (extended source) parameters, which are our basis for grouping the object types.

\begin{figure}
 \includegraphics[width=0.5\textwidth, angle=0]{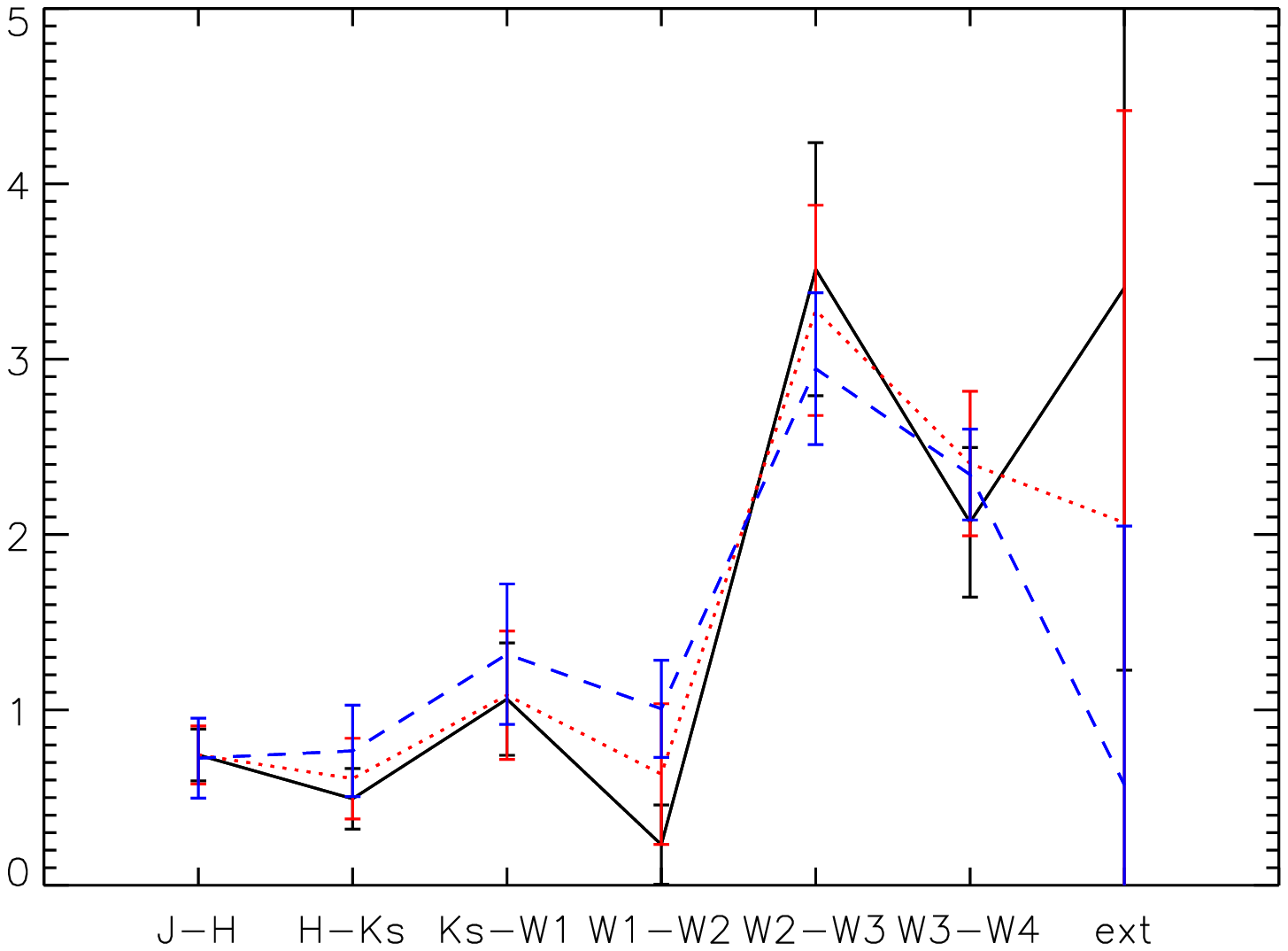}
\caption{Three main groups of SIMBAD extragalactic objects. Types put into G1 are shown with the black solid line. They appear to have lower near--IR colours and have mostly high \textit{ext} (extended source flag) values. Source types of G2 group are presented with blue dashed lines. They are less extended based on the \textit{ext} parameter value and have higher near--IR colours. Remaining sources are classified as G3 group and are presented with red dashed line.}
\label{exgal_ext}
\end{figure}

\begin{figure}
 \includegraphics[width=0.5\textwidth, angle=0]{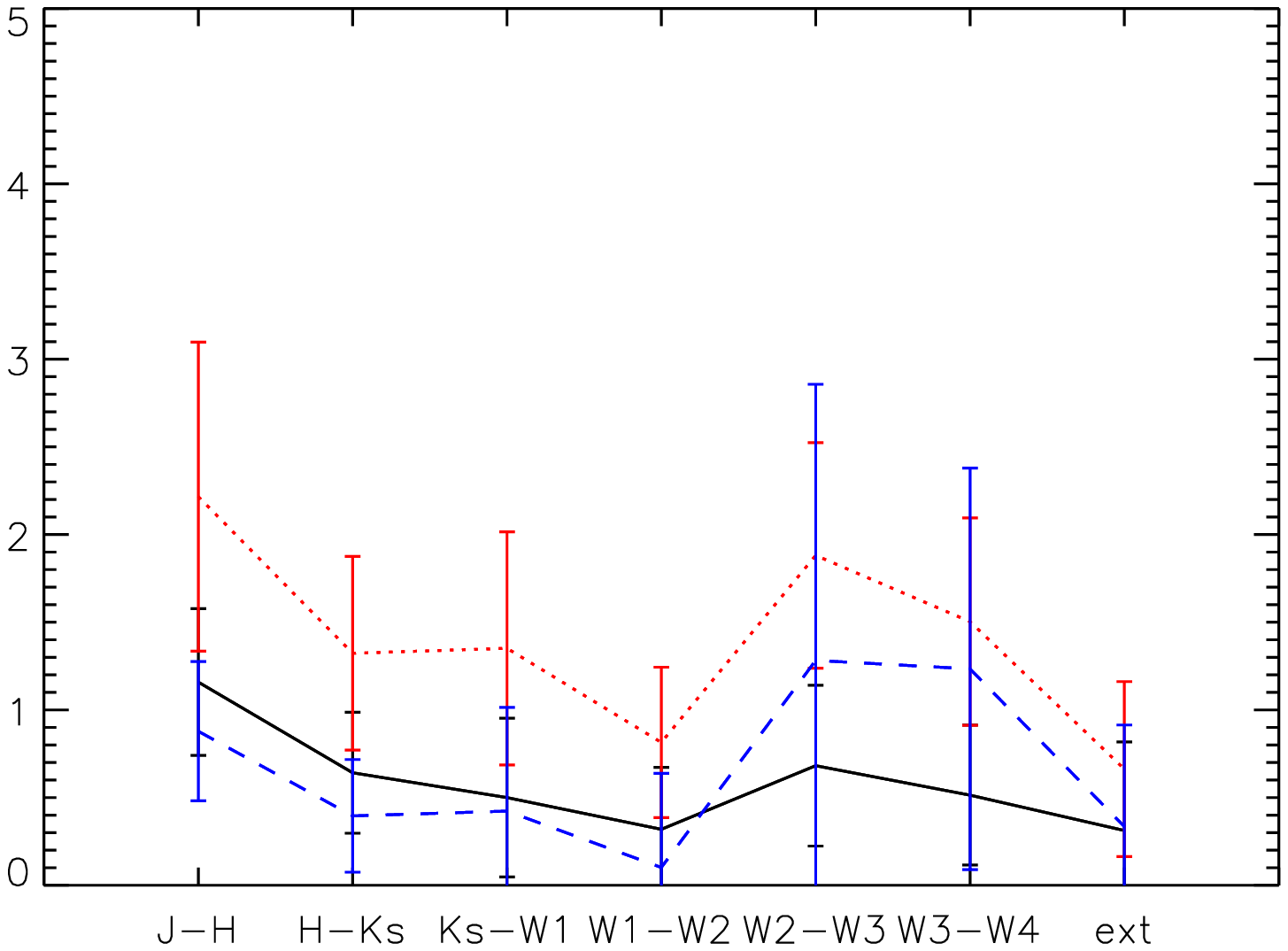}
\caption{Average colour indices and \textit{ext} values of three main groups of SIMBAD source types of evolved objects. Types put into E1 are shown with the black solid line. E2 types are presented with the red solid lines. They have high J-K colours compared to the other object types. E3 objects are shown with blue lines.}
\label{evo_ext}
\end{figure}

\begin{figure}
 \includegraphics[width=0.5\textwidth, angle=0]{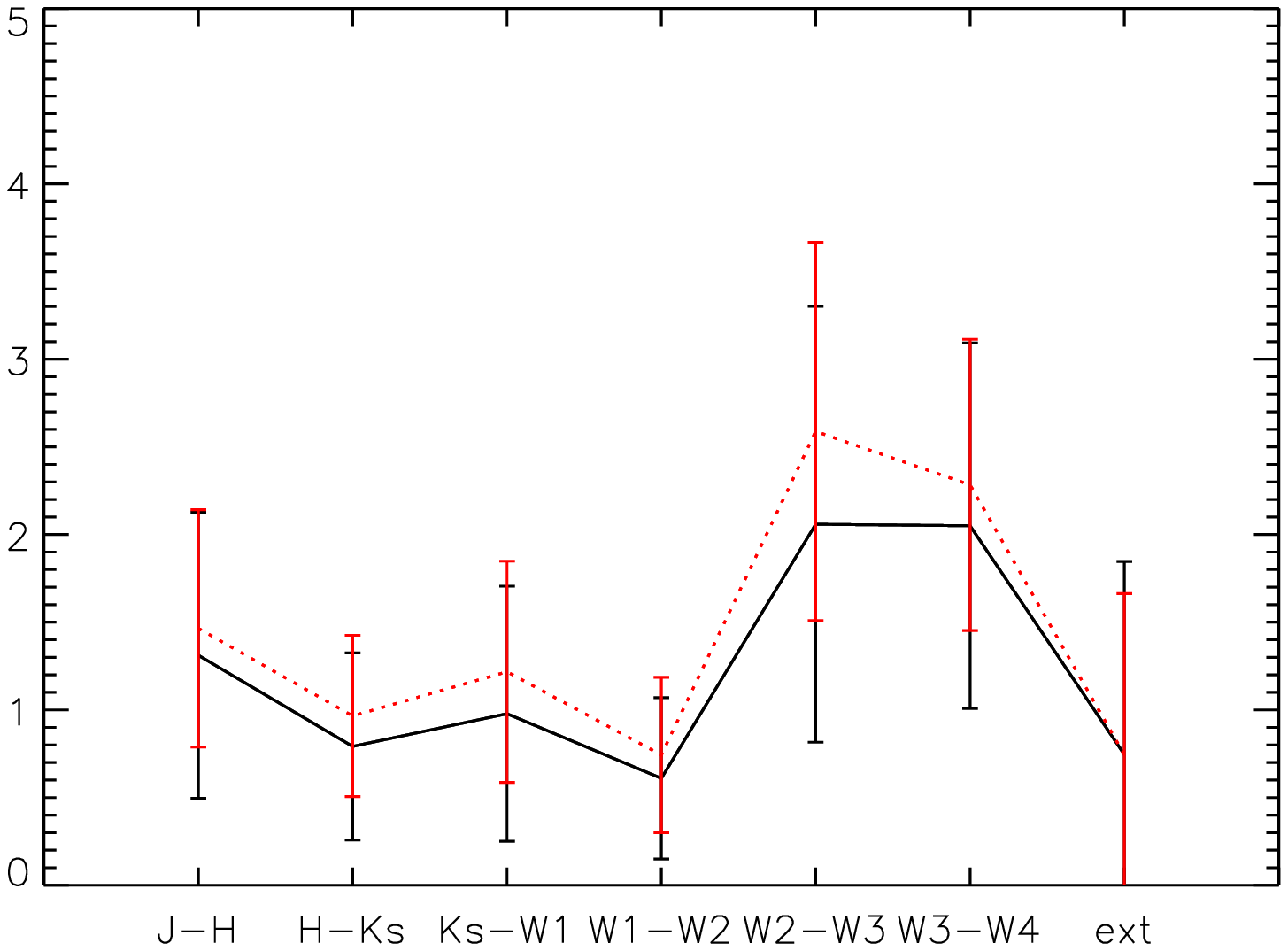}
\caption{Average colour indices and \textit{ext} values of two groups of SIMBAD source types of young objects. All average colour indices of Y1 sources (black solid line) are lower than that of Y2 objects (red dotted line).}
\label{yso_ext}
\end{figure}

\begin{figure}
 \includegraphics[width=0.5\textwidth]{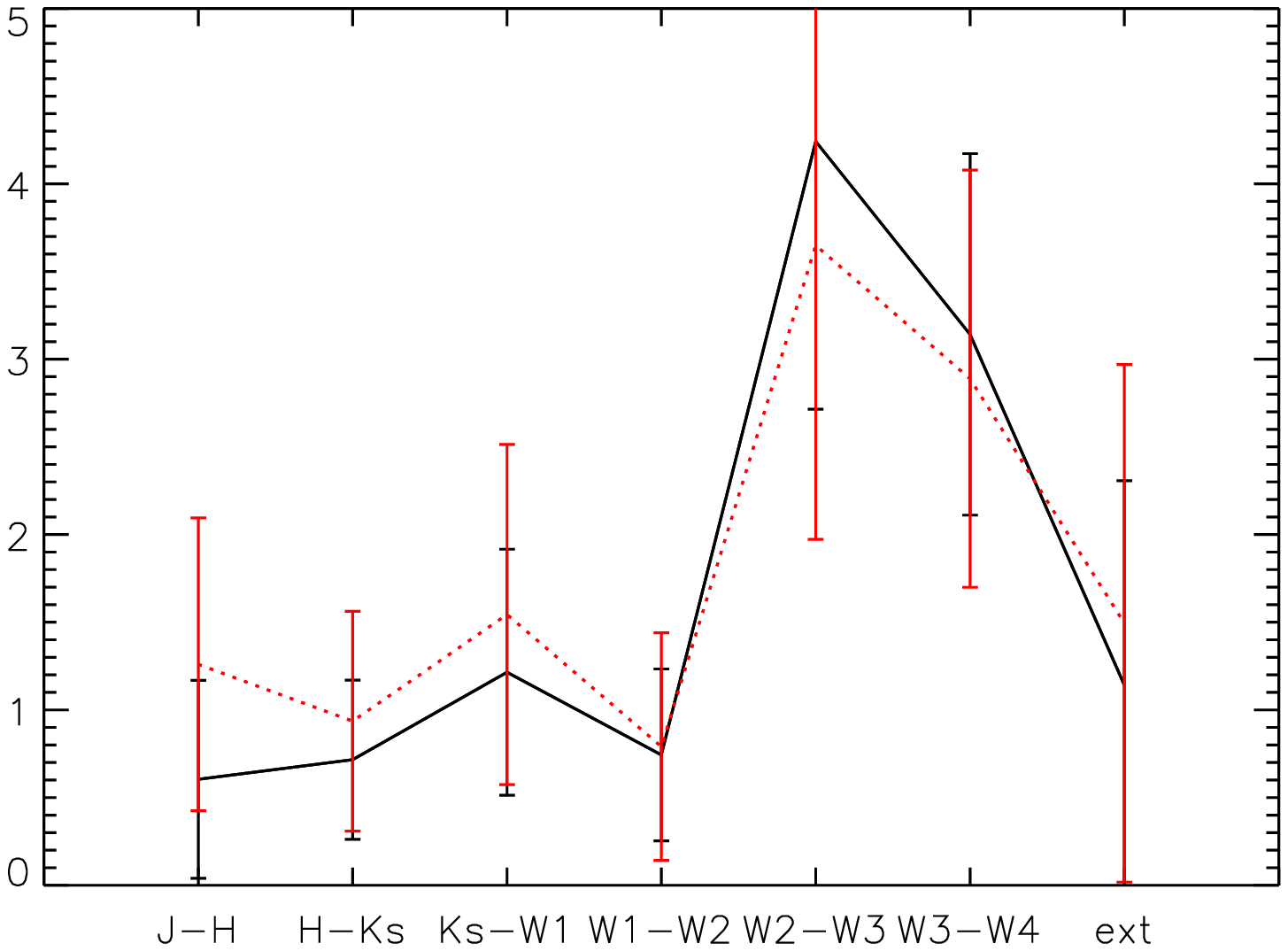}
\caption{Average colour indices and \textit{ext} values of two groups of SIMBAD source types of ISM related objects. ISM1 (black solid line) appear to be have lower near-IR colour indices that ISM2 (red dotted line).}
\label{ism_ext}
\end{figure}

\newpage
\section{Fraction of known YSOs as a function of magnitude in the WISE and 2MASS bands}
\begin{table*}
\begin{tabular}{cc}
 \includegraphics[width=0.35\textwidth,trim=0cm 0cm 0cm 0cm]{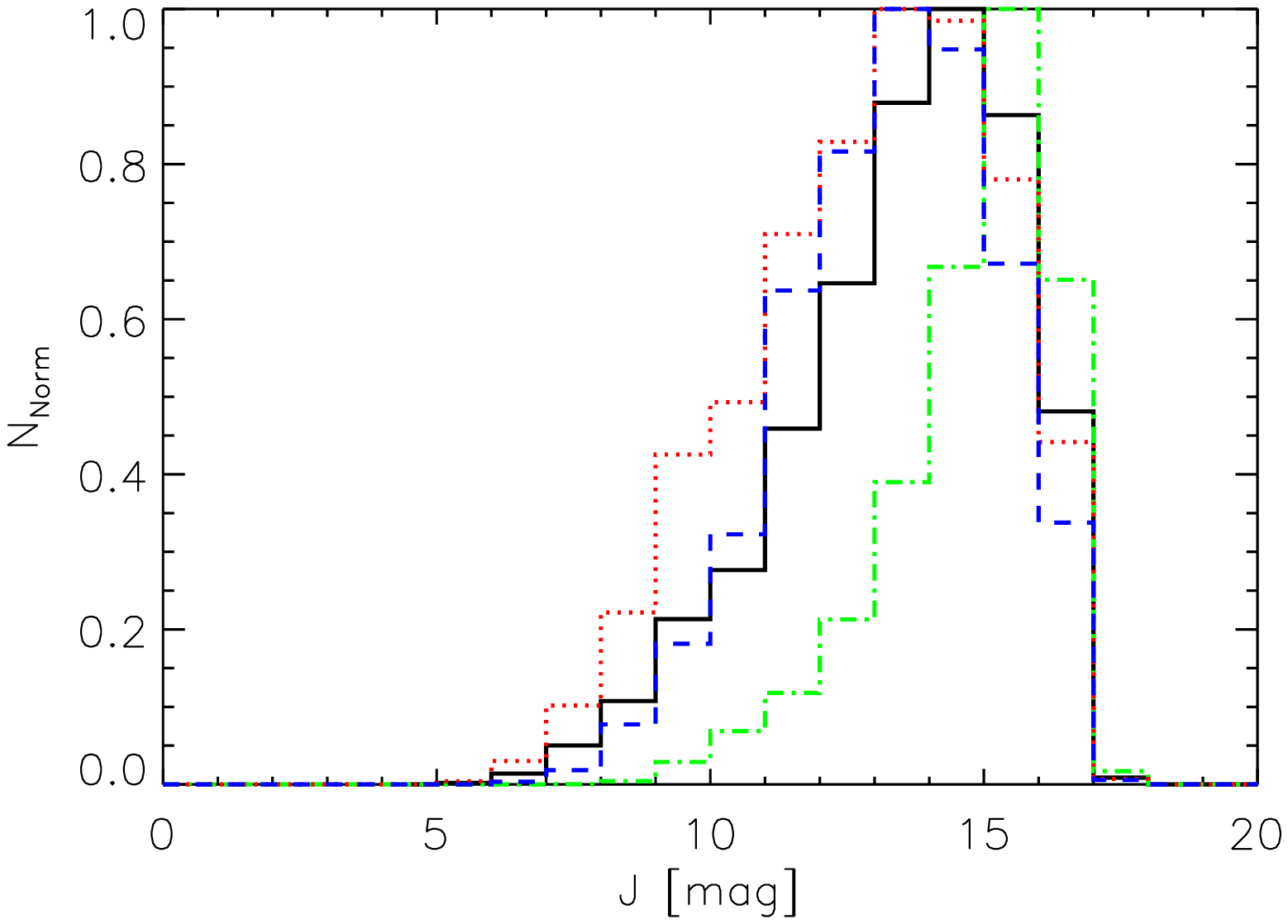} &  \includegraphics[width=0.35\textwidth,trim=0cm 0cm 0cm 0cm]{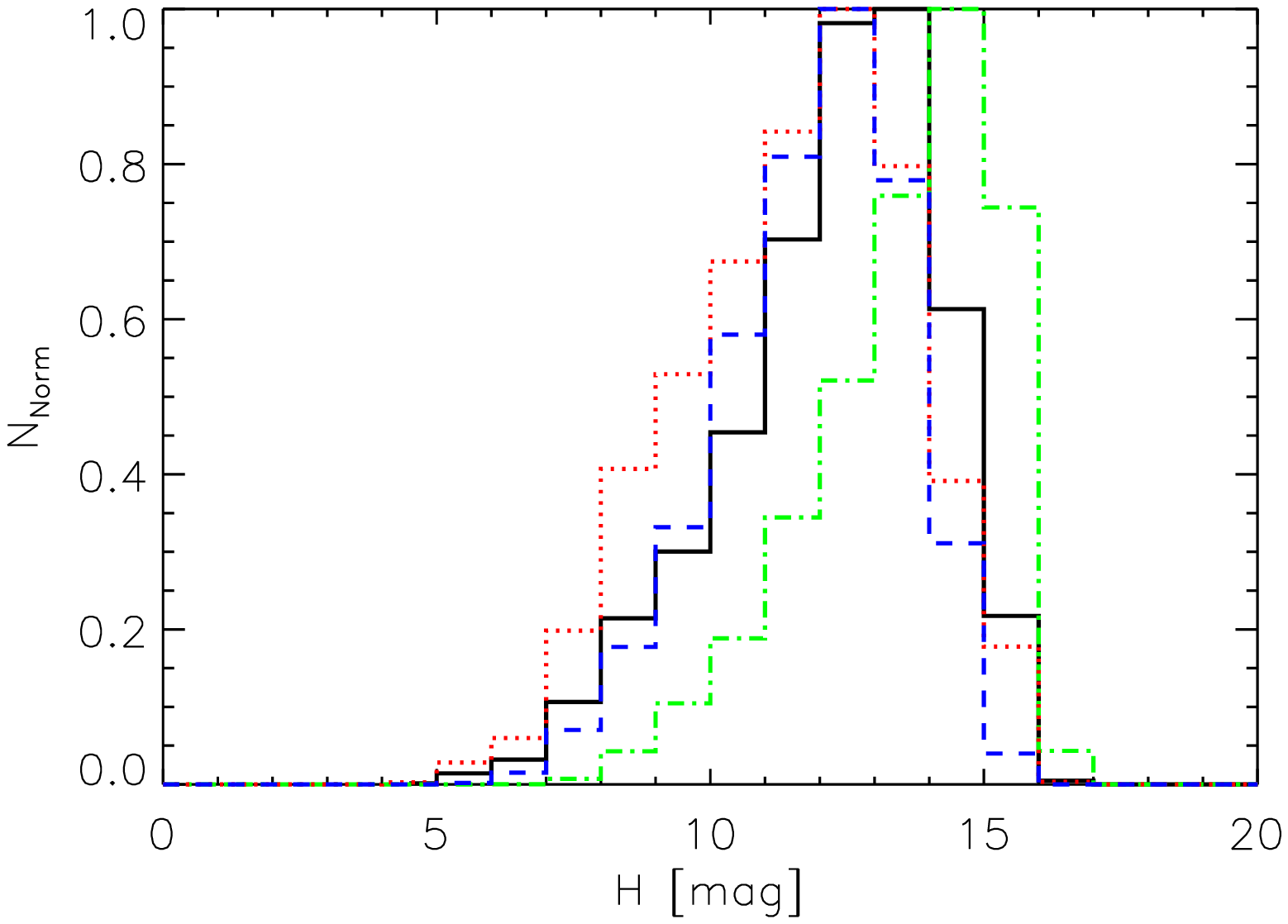}\\
 \\
 \includegraphics[width=0.35\textwidth,trim=0cm 0cm 0cm 0cm]{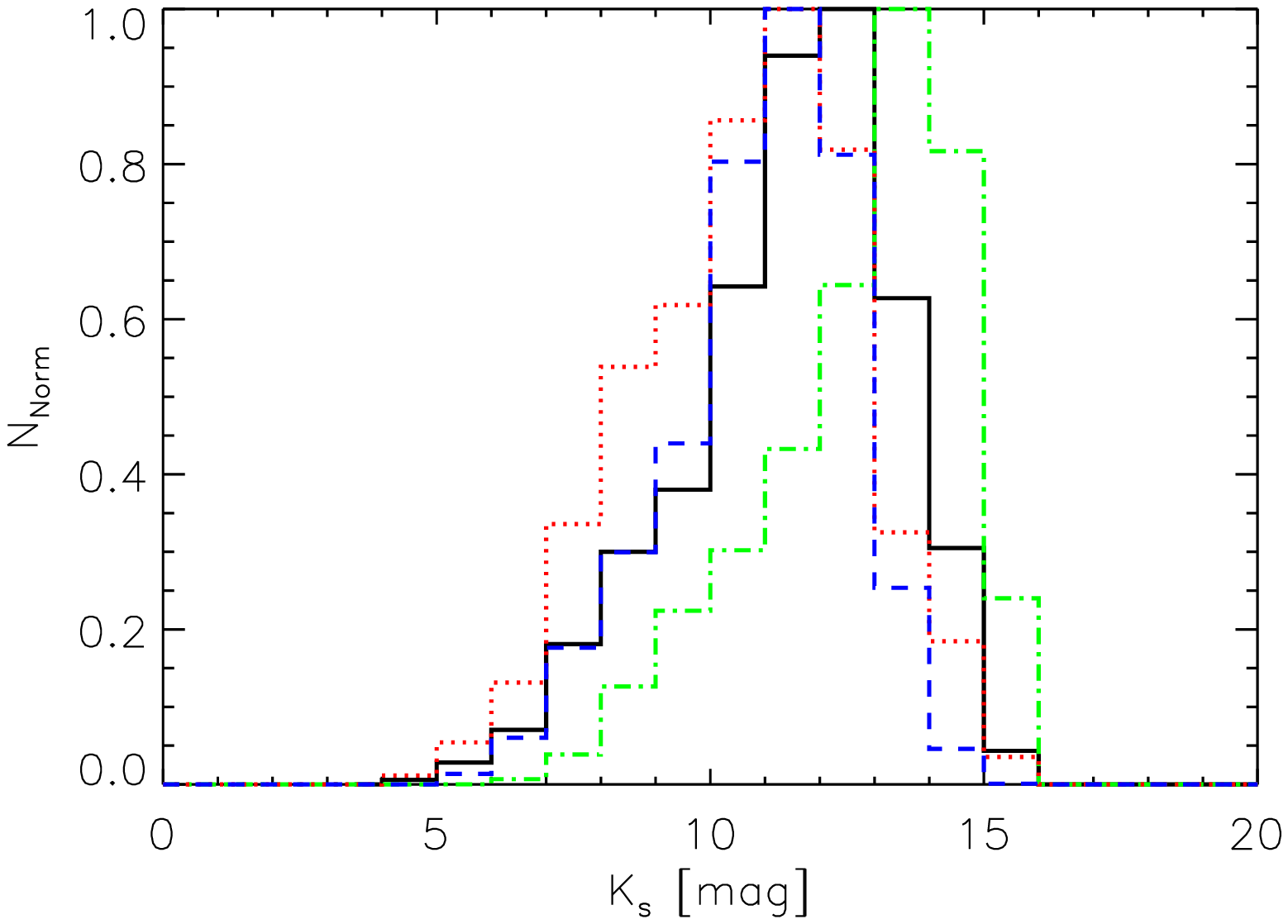} &  \includegraphics[width=0.35\textwidth,trim=0cm 0cm 0cm 0cm]{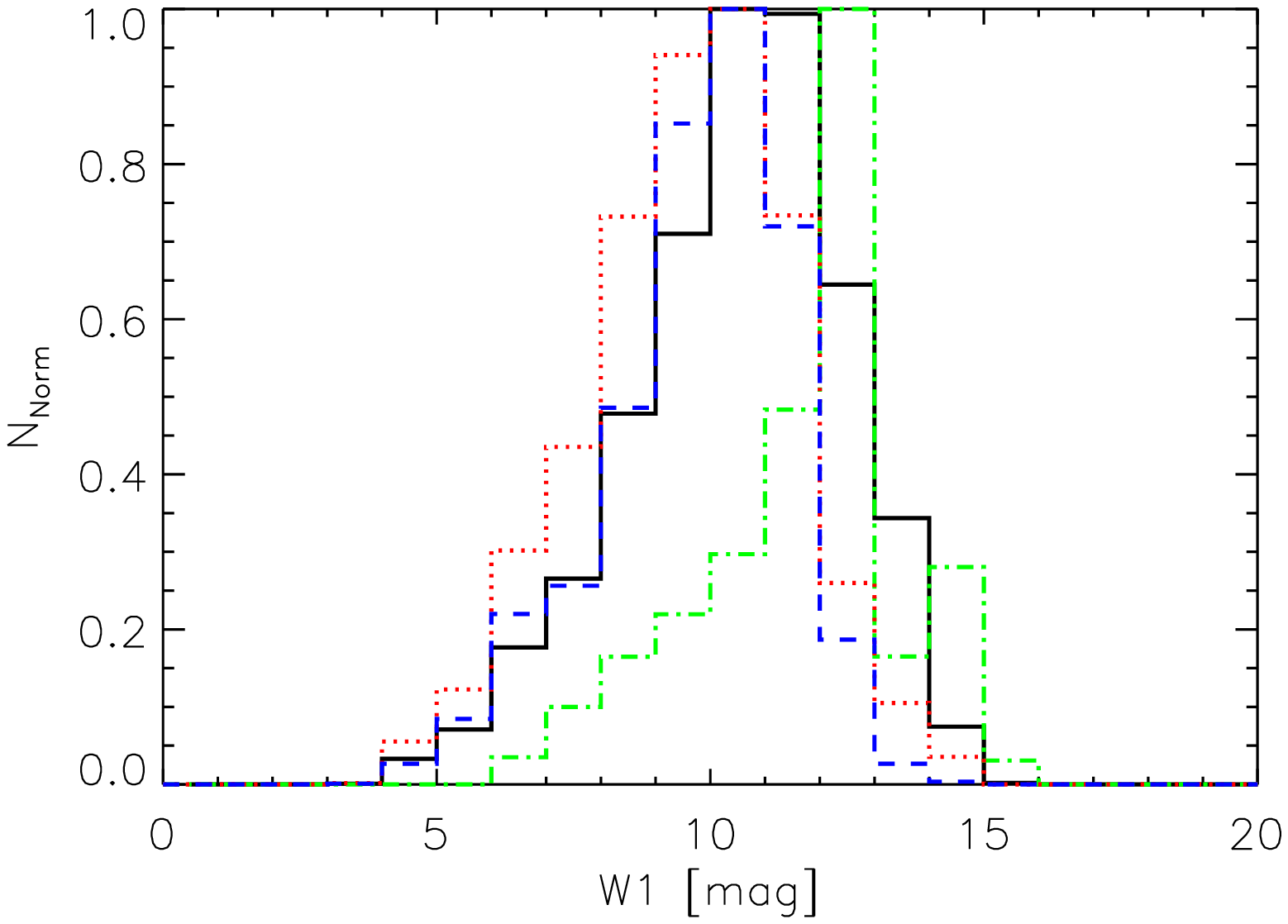}\\
 \\
 \includegraphics[width=0.35\textwidth,trim=0cm 0cm 0cm 0cm]{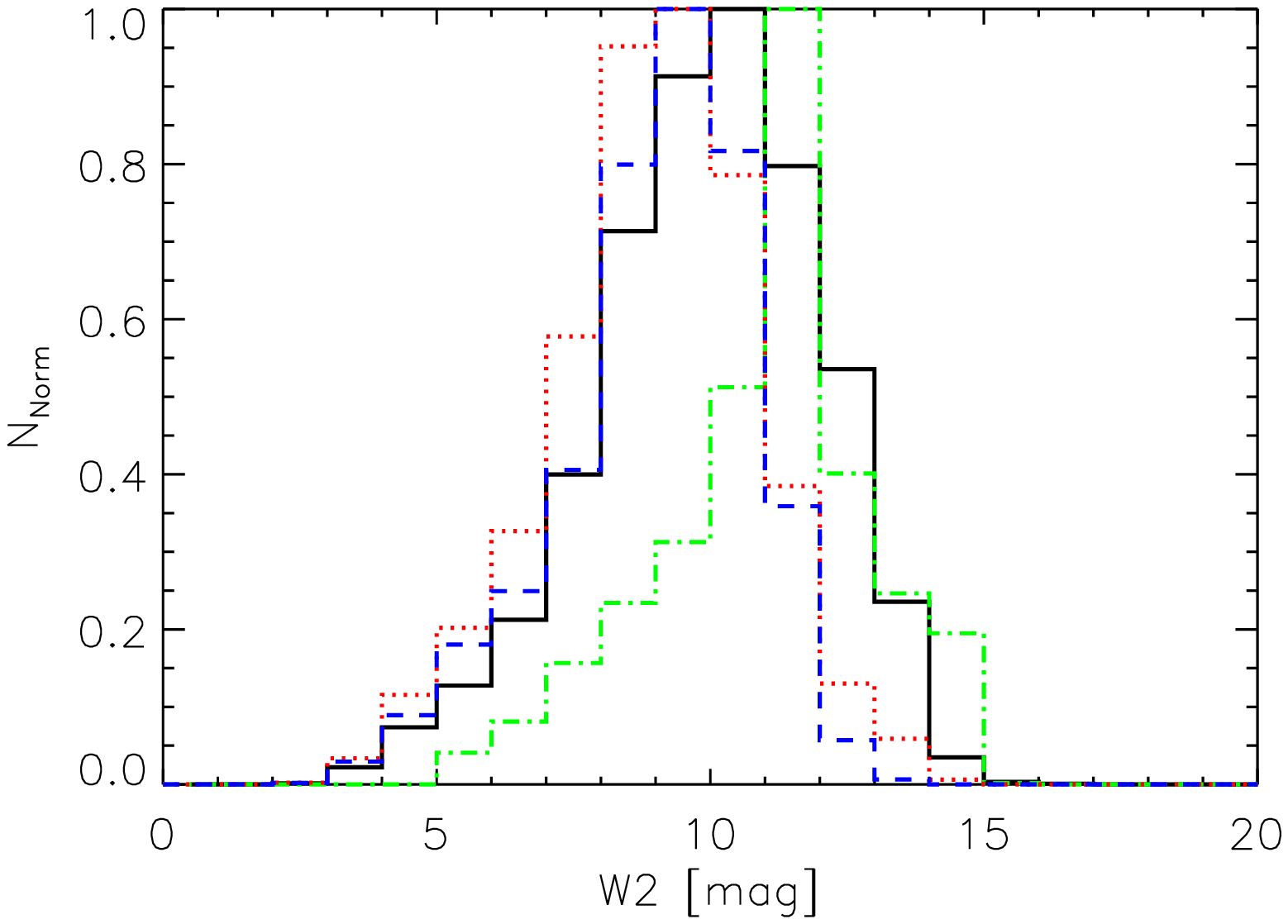} &  \includegraphics[width=0.35\textwidth,trim=0cm 0cm 0cm 0cm]{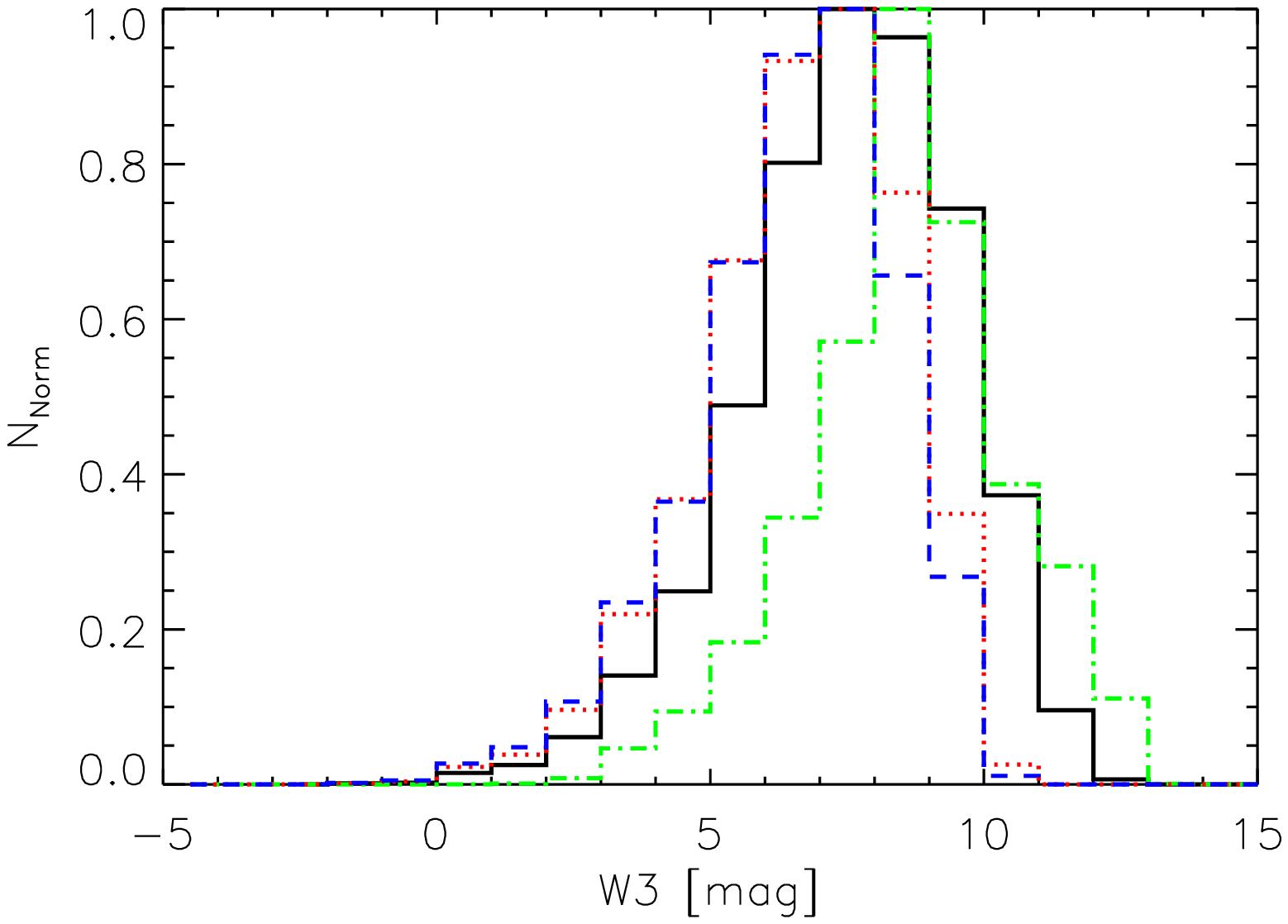}\\
 \\
 \includegraphics[width=0.35\textwidth,trim=0cm 0cm 0cm 0cm]{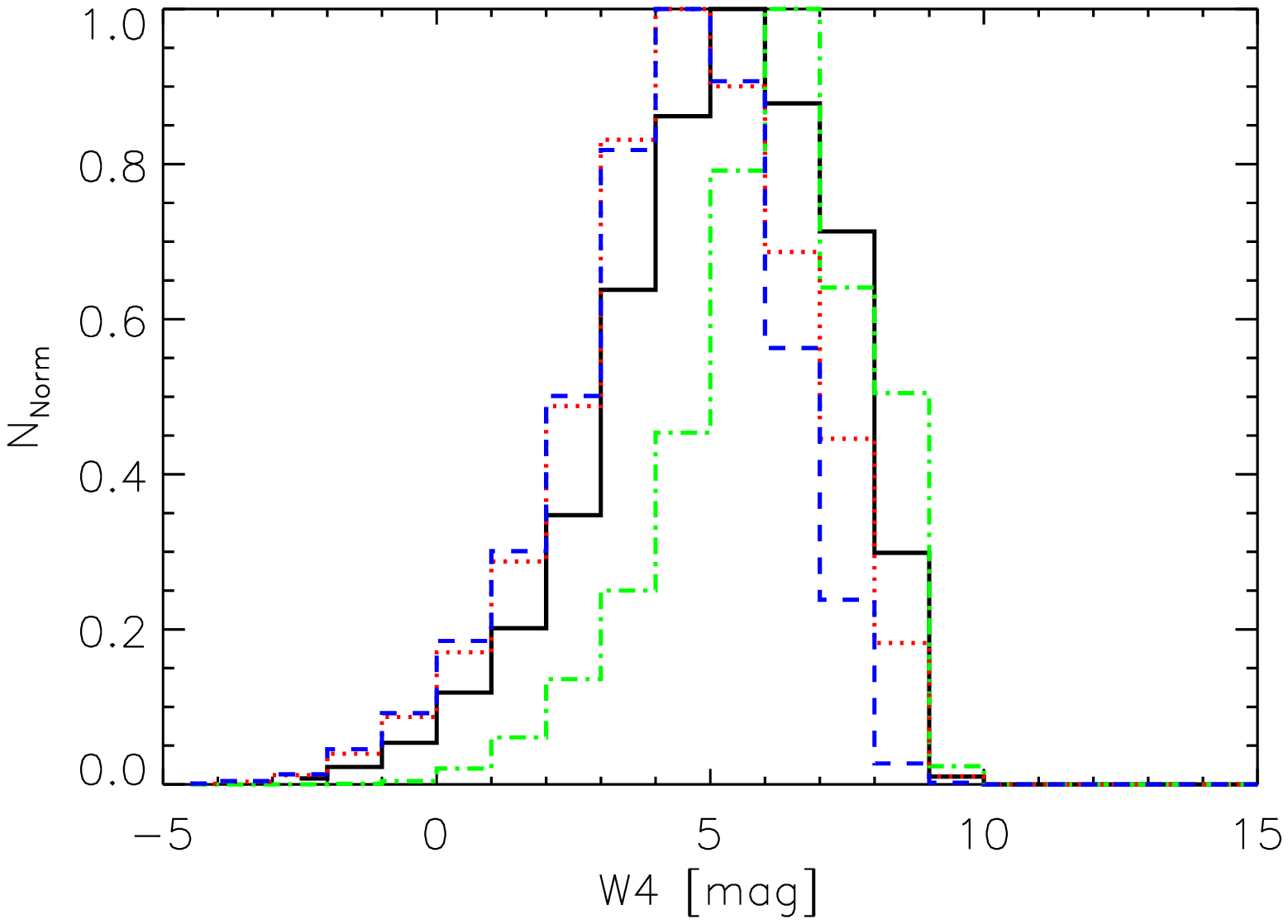} & \\
 
\end{tabular}
\caption{Fraction of known YSOs in the \textit{W0} (black solid line), \textit{real} (red dotted line), Class I/II (blue dashed line) and KL14 (green dashed dotted line) samples as a function of brightness in the different 2MASS and WISE bands.}\label{completenesshist}
\end{table*}

\bsp

\label{lastpage}

\end{document}